\pdfoutput=1
\documentclass[
    aps,pre,twocolumn,
    reprint,
    superscriptaddress,
    nofootinbib,
    floatfix,
    amssymb,
    longbibliography
]{revtex4-2}
\usepackage[utf8]{inputenc}
\usepackage{xcolor}
\usepackage{titlesec}
\usepackage{graphicx}
\usepackage[normalem]{ulem}
\usepackage{geometry}
\usepackage{newfloat}
\usepackage{amsmath}
\usepackage{lineno}
\usepackage[inline]{enumitem}  

\definecolor{linkColor}{rgb}{0,0.3,0.7}
\usepackage[colorlinks=true,
            allcolors=linkColor,
            pdfborder={0 0 0},
            pdfencoding = auto
]{hyperref}

\usepackage{textcomp,gensymb}

\begin{document}
\title{Light-induced cortical excitability reveals programmable shape dynamics in starfish oocytes}

\author{Jinghui Liu}
\thanks{These authors contributed equally.}
\affiliation{Department of Physics, Massachusetts Institute of Technology, Cambridge, MA, USA.}
\affiliation{Center for Systems Biology Dresden, Dresden, Germany.}

\author{Tom Burkart}
\thanks{These authors contributed equally.}
\affiliation{Arnold Sommerfeld Center for Theoretical Physics and Center for NanoScience, Department of Physics, Ludwig Maximilian University of Munich, Munich, Germany.}

\author{Alexander Ziepke}
\affiliation{Arnold Sommerfeld Center for Theoretical Physics and Center for NanoScience, Department of Physics, Ludwig Maximilian University of Munich, Munich, Germany.}

\author{John Reinhard}
\affiliation{Medical Biochemistry and Molecular Biology, Saarland University, Homburg, Germany.}
\affiliation{Physiology Course, Marine Biological Laboratory, Woods Hole, MA, USA.}

\author{Yu-Chen Chao}
\affiliation{Department of Physics, Massachusetts Institute of Technology, Cambridge, MA, USA.}
\affiliation{School of Engineering and Applied Sciences, Harvard University, Cambridge, MA, USA.}

\author{Tzer Han Tan}
\affiliation{Department of Physics, Massachusetts Institute of Technology, Cambridge, MA, USA.}
\affiliation{Current address: Department of Physics, University of California San Diego, La Jolla, CA, USA.}

\author{S.~Zachary Swartz}
\affiliation{Whitehead Institute for Biomedical Research, Cambridge, MA, USA.}
\affiliation{Eugene Bell Center for Regenerative Biology and Tissue Engineering, Marine Biological Laboratory, Woods Hole, MA, USA.}

\author{Erwin Frey}
\email{frey@lmu.de}
\affiliation{Arnold Sommerfeld Center for Theoretical Physics and Center for NanoScience, Department of Physics, Ludwig Maximilian University of Munich, Munich, Germany.}
\affiliation{Max Planck School Matter to Life, Munich, Germany.}

\author{Nikta Fakhri}
\email{fakhri@mit.edu}
\affiliation{Department of Physics, Massachusetts Institute of Technology, Cambridge, MA, USA.}

\date{September 08, 2024}

\begin{abstract}%
Chemo-mechanical waves on active deformable surfaces are a key component for many vital cellular functions.
In particular, these waves play a major role in force generation and long-range signal transmission in cells that dynamically change shape, as encountered during cell division or morphogenesis.
Reconstituting and controlling such chemically controlled cell deformations is a crucial but unsolved challenge for the development of synthetic cells.
Here, we develop an optogenetic method to elucidate the mechanism responsible for coordinating surface contraction waves that occur in oocytes of the starfish \textit{Patiria miniata} during meiotic cell division.
Using spatiotemporally-patterned light stimuli as a control input, we create chemo-mechanical cortical excitations that are decoupled from meiotic cues and drive diverse shape deformations ranging from local pinching to surface contraction waves and cell lysis.
We develop a quantitative model that entails the hierarchy of chemical and mechanical dynamics, which allows to relate the variety of mechanical responses to optogenetic stimuli. 
Our framework systematically predicts and explains transitions of programmed shape dynamics.
Finally, we qualitatively map the observed shape dynamics to elucidate how the versatility of intracellular protein dynamics can give rise to a broad range of mechanical phenomenologies.
More broadly, our results pave the way toward real-time control over dynamical deformations in living organisms and can advance the design of synthetic cells and life-like cellular functions.

\end{abstract}
\maketitle

The ability to dynamically control cell shape is a vital property of living organisms on a wide range of length scales, ranging from the coordination of cell division in yeast~\cite{Mishra.etal2012} and bacteria~\cite{Varma.etal2008} to morphogenesis in chicken~\cite{Shyer.etal2015} and fruit fly embryos~\cite{Lecuit.Lenne2007, Talia.Vergassola2022}.
In general, the geometric adaptation of an organism in response to cell-cycle and developmental cues is orchestrated by a complex network of chemical reaction circuits~\cite{Burkart.etal2022}.
When coupled to mechanical effectors such as the actomyosin cytoskeleton~\cite{Chalut.Paluch2016}, the chemo-mechanical pathway can generate contractile forces and cause dynamical cell shape changes that are crucial for geometric adaptation and remodeling of cells and tissues~\cite{Kabaso.etal2010, Martin.Goldstein2014, Dasbiswas.etal2018}.

Unveiling the physical principles that dictate the shape control mechanisms in cells is of great experimental and theoretical interest in a variety of emerging fields, including soft robotics~\cite{Majidi2019}, biomimetic material design~\cite{Grinthal.Aizenberg2013}, and synthetic cells~\cite{Mulla.etal2018,Schwille2019, McNamara.etal2023, Fu.etal2023}. 
However, the intricate nonlinearities of the underlying biochemical reaction networks and their coupling to guiding cues often render quantitative studies difficult.
One major challenge lies in the ability to isolate interesting components of chemo-mechanical circuits and gaining accurate spatiotemporal control over them \textit{in vivo}~\cite{Mcnamara.etal2016}.
\begin{figure*}[!t]
    \centering
	\includegraphics[width=\textwidth]{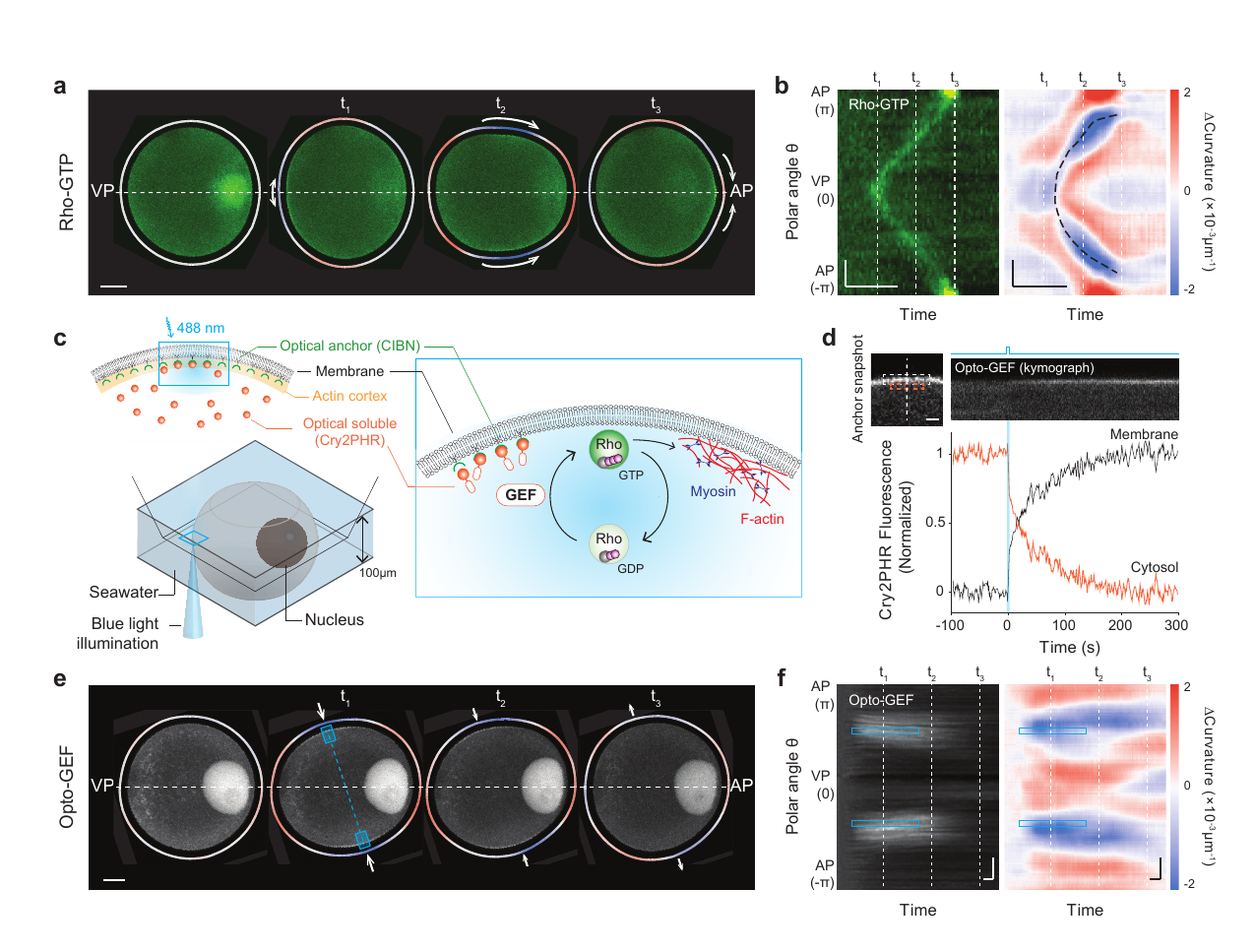}
        \caption{
        \textbf{Optogenetic GEF assay enables real-time control of cellular geometry in prophase-arrested starfish oocytes.}
        \textbf{a,} Midplane snapshots of a wild-type oocyte undergoing surface contraction wave (SCW) during meiotic anaphase. Fluorescence and ring overlay show active Rho-GTP density (labeled with rGBD-GFP) and cell surface curvature change, respectively. Scale bar: $50\,\mu \text{m}$.
        \textbf{b,} Kymographs of Rho-GTP fluorescence density and cell surface curvature during the SCW. Scale bar: $100\,\mu \text{m}$, 10~min. 
        \textbf{c,} Assay schematic of the optogenetic GEF (Opto-GEF) that reversibly translocates from cytosol to membrane upon blue light (488~nm) illumination. 
        \textbf{d,} Opto-GEF recruitment kinetics after a short pulse (1s) of regional illumination. Boxed regions in the snapshot of membrane anchor (CIBN-EGFP, upper left image) were used to quantify the cytosolic and membrane fractions of Opto-GEF shown in the graph. The kymograph on the right shows the Opto-GEF accumulation along the white dotted line. Scale bar (upper left): $5\mu \text{m}$. 
        \textbf{e,} Snapshots of a prophase-arrested photosensitive oocyte undergoing surface contractions driven by regional illumination. Location of the illuminated sites are shown as cyan boxes. Scale bar: $50\,\mu \text{m}$. 
        \textbf{f,} Kymographs of Opto-GEF fluorescence density and cell surface curvature during and after illumination. Scale bar: $100\,\mu \text{m}$, 10~min.
        }
 	\label{fig1} 
\end{figure*}

We address this challenge by designing an optogenetic switch that exerts quantitative controls on the cortical activities induced by the GTPase Rho in prophase-arrested oocytes of the starfish \textit{Patiria miniata}.
Recent studies have reported chemo-mechanical \emph{surface contraction waves} (SCWs) traveling across starfish oocytes during meiotic anaphase~\cite{Bischof.etal2017, Bement.etal2015,Tan.etal2020,Wigbers.etal2021} (Fig.~\ref{fig1}a-b, Extended Data Fig.~\ref{ExtFig1}a-b). 
These waves are primarily triggered by an enhanced actomyosin contractility~\cite{Foster.etal2022} stimulated by high concentrations of GTP-bound Rho near the cell surface~\cite{Narumiya.etal2009}.
At the heart of these Rho-GTP concentration peaks lies a nonlinear biochemical interaction between Rho and GEF proteins~\cite{Wigbers.etal2021}, as GEF enzymes promote the direct and autocatalytic conversion of Rho-GDP to Rho-GTP~\cite{Rossman.etal2005, Bement.etal2024}.
The spatio-temporal GEF dynamics are therefore the key to controlling the actomyosin activity in starfish oocytes and a possible gateway to introduce optogenetic stimuli to the chemo-mechanical system.

Here, we explore the programmable shape dynamics attained by optogenetic control over the intracellular protein dynamics that are decoupled from cell-cycle cues.
By designing an optogenetic assay in prophase-arrested starfish oocytes, we achieve localized real-time modulation of Rho cortical excitability and cell shape control.
The optogenetic control of RhoGEFs for inducing mechanical responses has been established across model organisms over the past decade, including the remodeling of epithelial tissue~\cite{Cavanaugh.etal2020, Varadarajan.etal2022, Mahlandt.etal2023}, migration of HeLa~\cite{Wu.etal2009, Wagner.Glotzer2016, Mahlandt.etal2023} and NIH3T3 cells~\cite{Levskaya.etal2009, Wagner.Glotzer2016}, and cell contractions of U2OS cells~\cite{Kamps.etal2020}.
However, comprehensive models that allow tracing the signal processing cascade from light activation to cell deformation have not been developed so far.
Here, we combine models for light-induced activation of optogenetic GEFs, Rho dynamics, and contraction-induced cell deformations to identify two key mechanisms of chemo-mechanical signal processing in the oocytes (based on diffusive fluxes and dynamic local reactive equilibria, respectively).
Our theoretical model elucidates how these mechanisms can be triggered selectively using light stimuli, thereby leading to diverse dynamical shapes, and allows to predict the mechanical response to arbitrary photoactivation cues.
We finally present a qualitative map that condenses the experimental and theoretical findings by relating the observed shape phenotypes to physical properties of the chemical reaction-diffusion system.
Our map is applicable to generic chemo-mechanical systems and can be used to navigate the parameter space when developing shape control mechanisms for other organisms.

\section{Optical control of GEF dynamics}%
In starfish oocytes, the Rho nucleotide exchange leading to enhanced actomyosin activity is primarily catalyzed by GEF activity on the membrane~\cite{Rossman.etal2005}.
Thus, one way to gain control over the actomyosin-regulated shape dynamics in oocytes decoupled from cell-cycle cues is to externally manipulate the spatio-temporal distribution of GEF, and in particular its membrane localization.
To this end, we designed an optogenetic switch that conjugates the catalytic DH domain of the endogenous GEF enzyme, Ect2, with a photosensitive protein domain (CRY2PHR)~\cite{Kennedy.etal2010}. 
Upon light stimuli, this photosensitive tag complex (Opto-GEF) can reversibly bind to its partner domain (CIBN) that is conjugated with a plasma membrane anchor (Fig.~\ref{fig1}c, Methods).
Starfish oocytes expressing both switch components showed a rapid (half-time ${\tau_\text{on}\approx 10.5 \, \text{s}}$) membrane recruitment of CRY2PHR proteins from the cytosol upon a short pulse (${\sim 1\,\text{s}}$) of blue light (488~nm) illumination (Fig.~\ref{fig1}d, Extended Data Fig.~\ref{ExtFig2}a-c).
This recruitment is reversible upon switching off the light with slow unbinding dynamics (half-time ${\tau_\text{off}\approx 7 \, \text{min}}$, Extended Data Fig.~\ref{ExtFig2}d-f), consistent with previously reported CRY2-CIBN interaction kinetics~\cite{Taslimi.etal2016}.
By design, the optogenetic constructs do not contain a GEF PH domain to minimise membrane association in the absence of light activation.

To validate that Opto-GEF interacts sufficiently strong with Rho to drive actomyosin contractions similar to those observed for wild-type meiotic oocytes (Extended Data Fig.~\ref{ExtFig1}c-d), we illuminated a small region close to the oocyte membrane (\emph{regional illumination}, $15 \times 30 \, \mu \text{m}^2$; Fig.~\ref{fig1}e), and monitored changes in the Opto-GEF density on the membrane and the correlated oocyte shape deformations (Fig.~\ref{fig1}f).
Indeed, localized mechanical responses were triggered shortly after the Opto-GEF accumulation, and deformed oocytes relaxed to their original shape after the light was switched off (Fig.~\ref{fig1}f, Extended Data Fig.~\ref{ExtFig1}e-f, Movie S1).
By expressing a Rho-GTP probe alongside optogenetic vectors with a GFP-free membrane anchor (Methods), we verified that the probed Rho-GTP membrane density closely follows the kinetics of Opto-GEF membrane accumulation (Extended Data Fig.~\ref{ExtFig3}a,b,d, Movie S2-S3).
To confirm that the cortex deformation is a result of enhanced actomyosin contractility, we also imaged the myosin dynamics following light activation by expressing a GFP-tagged component of the non-muscle myosin II alongside the optogenetic vectors (Methods).
We found that after illumination, myosin II is recruited to near-membrane regions where Opto-GEF also accumulates (Extended Data Fig.~\ref{ExtFig4}a-d).
Taken together, we conclude that the Opto-GEF switch attains control over actomyosin activity and the cell shape by modulating the Rho GTPase cycling kinetics.

\section{Translating optical stimuli into chemo-mechanical responses}%
\begin{figure*}[!t]
\centering
	\includegraphics[width=1\textwidth]{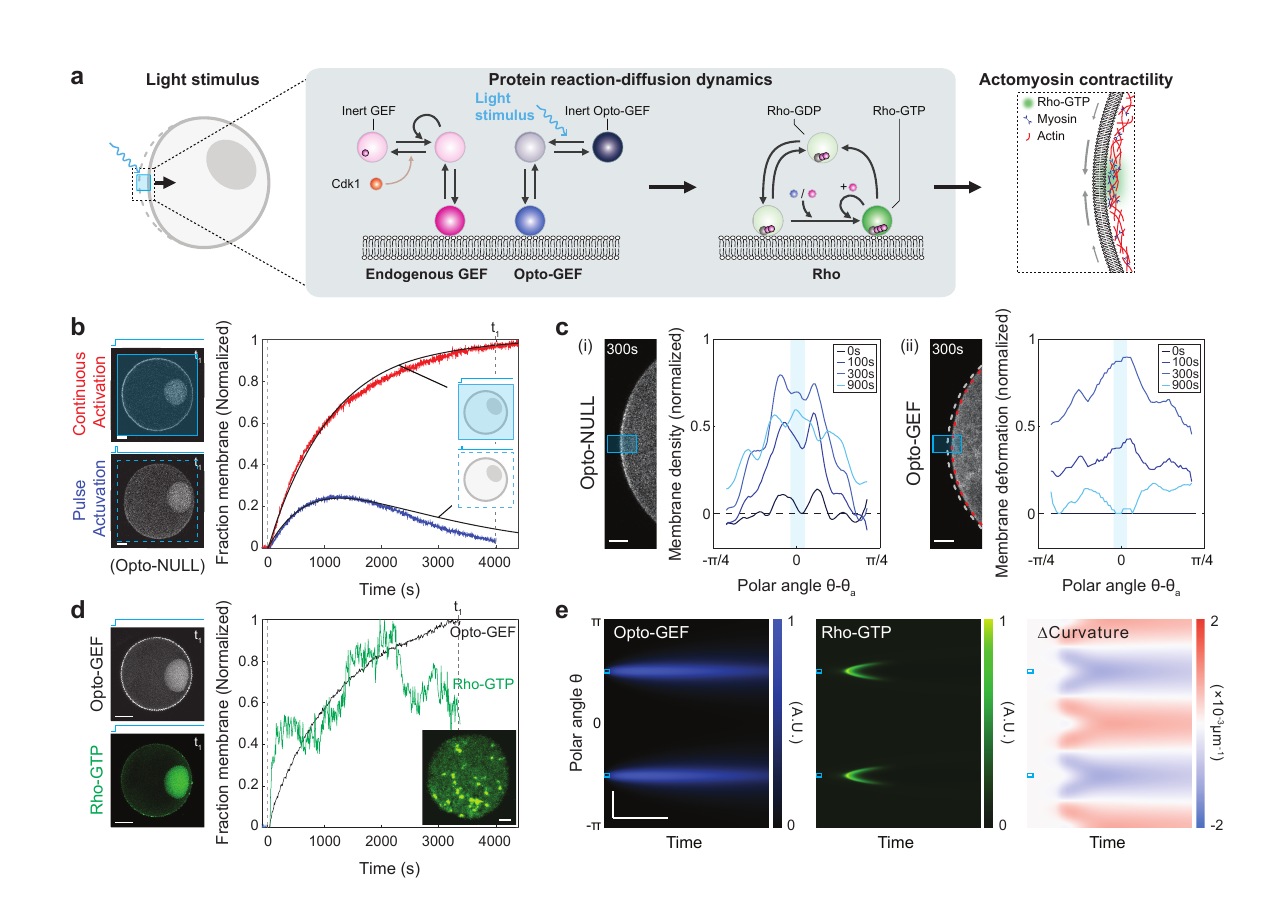}
	\caption{\textbf{A hierarchically-coupled chemo-mechanical model quantitatively captures the dynamical cellular geometry in response to versatile optogenetic control inputs.}
        \textbf{a,} Schematic of the chemo-mechanical model (Methods, full description in Supplementary Information).
        Inert Opto-GEF (purple) is activated by a light stimulus (blue), enabling it to bind to the membrane. On the membrane, both the endogenous GEF (pink) and Opto-GEF mediate Rho nucleotide exchange (green).
        The cell is considered as a deformable surface where contraction is driven by Rho-GTP accumulation.
        \textbf{b,} Membrane recruitment kinetics of the photosensitive protein tag complex after pulse (blue) and continuous (red) global illumination. Results are obtained from experimental measurements from Opto-Null assay and model simulation (black line). Scale bar: $50\,\mu \text{m}$. 
        \textbf{c,} Under the same regional pulse (1s) illumination, oocytes expressing the Opto-Null assay (i, snapshot on the left) exhibit membrane recruitment (i, graph on the right) without mechanical responses, while oocytes expressing the functional Opto-GEF assay (ii, snapshot on the left) exhibit both membrane recruitment (data not shown in this plot) and mechanical deformation responses (ii, graph on the right). Graphs in (i) and (ii) show the normalized recruitment and deformation responses near the illumination site on membrane. Dashed lines in both graphs show the basal reference state taken in the normalization (Methods). The Opto-Null recruitment and Opto-GEF deformation responses peak similarly around t=300~s post illumination. Extended Data Fig.~\ref{ExtFig7}a-b shows full quantification of this experiment. Scale bar: $20\,\mu \text{m}$. 
        \textbf{d,} Membrane recruitment kinetics of Opto-GEF and Rho-GTP after continuous global illumination. Inset shows a snapshot of Rho-GTP accumulation near membrane at the end of the continuous global illumination (See also Movie S3). Scale bar: $50\,\mu \text{m}$. 
        \textbf{e,} Simulated kymographs of Opto-GEF, Rho-GTP and cell surface curvature during surface contractions driven by regional illumination. Parameters are stated in the Supplementary Information. Scale bar: $100\,\mu \text{m}$, 10~min. 
	}
	\label{fig2}
\end{figure*}%
We are interested in using the optical control of spatio-temporal GEF dynamics to realize more complex shape dynamics that can be predicted based on the nonlinear biochemical interaction between Rho and GEF proteins.
To this end, we extended a previously developed model for the Cdk1/Ect2/Rho reaction-diffusion dynamics by an optogenetic GEF component~\cite{Wigbers.etal2021} (Fig.~\ref{fig2}a).
For this, RhoGEFs are assumed to contribute to the autocatalytic activation of Rho-GTP in addition to mediating the direct nucleotide exchange~\cite{Bement.etal2015, Goryachev.etal2016, Wigbers.etal2021, Michaud.etal2022}.
We note that the mechanisms underlying the autocatalytic activation are not yet fully understood~\cite{Michaud.etal2022, Bement.etal2024}, though recently a structural regulation mechanism has been reported for the human variant of endogenous GEF (Ect2)~\cite{Chen.etal2020}.
In the model, the endogenous autocatalytic interaction is implemented merely as an effective term (Supplementary Information).
No autocatalytic activation via Opto-GEF is expected because the PH domain essential for Ect2-dependent autocatalytic Rho activation~\cite{Chen.etal2020} is cleaved.
We furthermore explicitly coupled the Rho-GTP dynamics to the cortex mechanics to incorporate deformations of the actomyosin cortex and contractile cortical flows.
Specifically, we modeled the oocyte surface (membrane and cortex) as an elastic contractile material based on a free energy functional involving cortical tension~\cite{Dubus.Fournier2006} and Canham-Helfrich bending~\cite{Canham1970}, where Rho-GTP is assumed to directly cause surface contractions~\cite{Narumiya.etal2009, Yin.etal2021} (Methods, Supplementary Information).
In addition, we used an active stress model to account for in-plane contractions of the oocyte cortex that lead to advective flows of membrane-bound proteins~\cite{Nishikawa.etal2017} (Supplementary Information).

Combining the biochemical module of the GEF-Rho reaction--diffusion dynamics with the mechanical module of the cell cortex contractions, we arrived at a comprehensive chemo-mechanical model for the optogenetically-modulated cell shape dynamics.
We validated the mathematical model and fixed parameters using two distinct experimental assays: Pre-meiotic oocytes expressing a photosensitive protein tag complex lacking the catalytic GEF domain (Opto-Null, for gauging the binding kinetics of the photosensitive GEFs; Fig.~\ref{fig2}b,c, Extended Data Fig.~\ref{ExtFig3}a,c, Methods), and wild-type oocytes undergoing SCWs where the Rho-GTP density and the oocyte geometry are tracked explicitly (for gauging the mechanical deformation; Fig.~\ref{fig2}d,e, Extended Data Fig.~\ref{ExtFig1}c-f, Methods).
This model accurately reproduced the light-induced deformations on the local scale (i.e., membrane curvature, Fig.~\ref{fig1}b,f) and on the scale of the entire oocyte (i.e., amplitudes of Fourier modes, Fig.~\ref{fig2}e, Supplementary Fig.~S7, Supplementary Information).

\section{Guided waves and trigger waves}%
\begin{figure*}[!t]
\centering
	\includegraphics[width=1\textwidth]{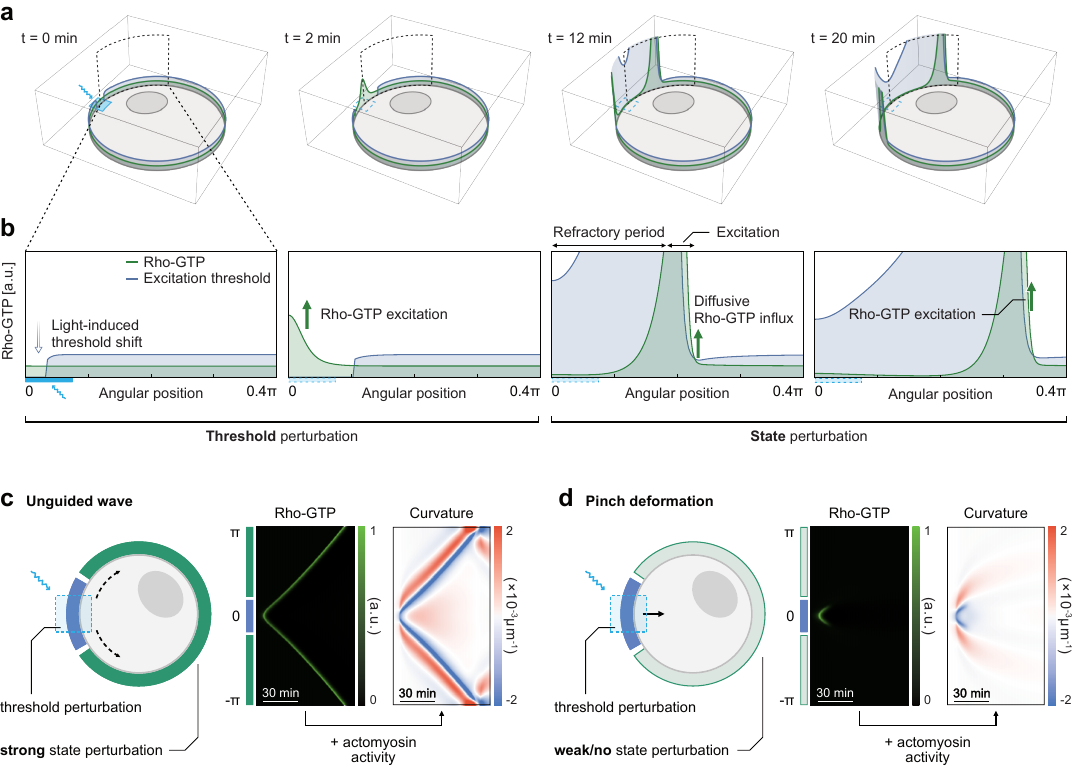}
	\caption{\textbf{Rho-GTP excitations can be triggered by state perturbations and threshold perturbations.}
            \textbf{a,}~Simulation of the Rho-GTP concentration (green) and excitation threshold (dark blue). A regional illumination (cyan square) causes a localized Rho excitation (left) which then spreads out into a traveling Rho wave (right).
            \textbf{b,}~Detailed view of the regions indicated by a black dotted frame in (\textbf{a}). A Rho excitation is triggered when the excitation threshold is brought below the Rho-GTP concentration (left, threshold perturbation by illumination) or when the Rho-GTP concentration is brought above the threshold (right, state perturbation by diffusive Rho-GTP fluxes on the membrane).
            \textbf{c,}~A regional light stimulus can trigger a self-sustained Rho-GTP wave that travels along the circumference of the oocyte for a sufficiently strong state perturbation.
            \textbf{d,}~For weak or subthreshold state perturbations, this wave decays quickly, yielding pinch-like deformations.%
	}
	\label{fig3}
\end{figure*}%
\begin{figure*}[!t]
    \centering
	\includegraphics[width=1\textwidth]{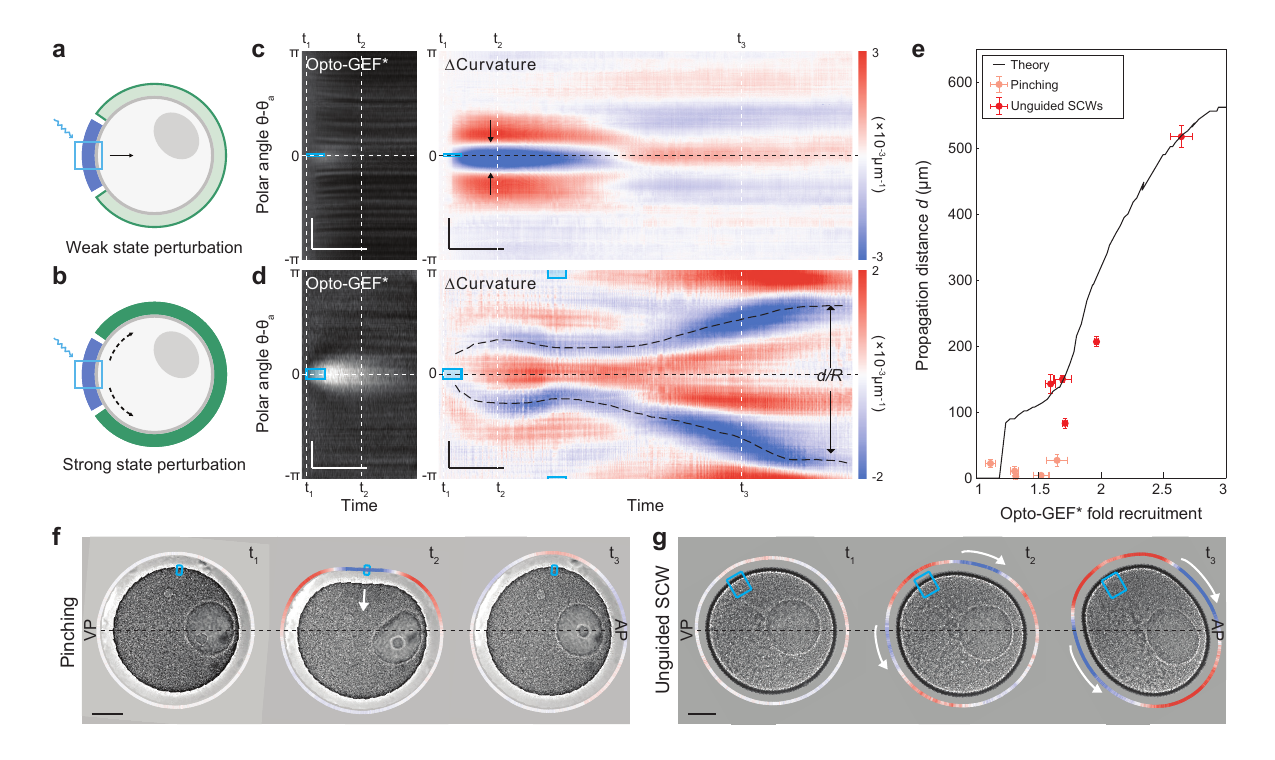}
	\caption{\textbf{Pinching and unguided surface contraction waves (SCWs) are experimentally realized with regional illumination assays.}
        \textbf{a,}~Schematic arrangement of threshold (blue) and state (green) perturbations leading to a pinching deformation or 
        \textbf{b,}~to an unguided wave.         
        \textbf{c,}~Membrane Opto-GEF* intensity (left) and curvature (right) kymographs corresponding to the experimental realization of a pinching oocyte. Regional illumination (cyan boxes) was turned on for 200~s. Scale bar: $100\,\mu \text{m}$, 10~min.
        \textbf{d,}~Membrane Opto-GEF* intensity (left) and curvature (right) kymographs corresponding to the experimental realization of an unguided SCW. Regional illumination (cyan boxes) was turned on for 200~s. Scale bar: $100\,\mu \text{m}$, 10~min.
        \textbf{e,} Experimental measurements of propagation distances for unguided SCWs as a function of regional Opto-GEF* recruitment strengths show qualitative agreement with model prediction. Importantly, the transition between pinching and unguided SCWs in oocytes is captured. Error bars represent statistical errors for Opto-GEF* intensity and propagation distances extracted from individual oocyte (see Methods for details).
        \textbf{f,} Representative timelapse snapshots of the pinching oocyte (same as (\textbf{c})). Colored rings show the instantaneous distribution of curvature along membrane. Scale bar: $50\,\mu \text{m}$. 
        \textbf{g,} Representative timelapse snapshots of the oocyte exhibiting unguided SCW (same as (\textbf{d})). Colored rings show the instantaneous distribution of curvature along membrane. Scale bar: $50\,\mu \text{m}$.
        }
	\label{fig4}
\end{figure*}%
Predicting the oocytes' shape response to optical stimuli requires a mechanistic understanding of the Rho dynamics, particularly the generation of traveling Rho-GTP peaks.
The Rho system is an excitable system, where an excitation is characterized by a sudden localized increase in the Rho-GTP concentration followed by a relaxation towards a stable steady state~\cite{Meron1992, Desai.Kapral2009} (refractory period).
Such an excitation only takes place if the Rho-GTP concentration exceeds a certain threshold concentration~\cite{Meron1992, Cross.Hohenberg1993, Desai.Kapral2009} (\emph{excitation threshold}).
The value of the excitation threshold depends on the reaction rates in the Rho system and therefore on the local GEF concentration, among others (Methods, Extended Data Fig.~\ref{ExtFig5}).
In the following, we provide a heuristic discussion of the main aspects for cell-cycle or optically-dependent Rho excitability (comprehensive analysis in Supplementary Information).

In general, two qualitatively different mechanisms can trigger a Rho excitation.
First, the excitation threshold can be lowered by rapidly increasing the membrane GEF concentration either following the removal of Cdk1 inhibition or, as in our case, by using localized illumination (Fig.~\ref{fig3}a,b, $t = 0\text{ min}$; Movie S4).
As soon as the threshold (blue) is lowered beneath the current Rho-GTP concentration (green), a Rho activity burst is triggered ($t = 2\text{ min}$).
We refer to this mechanism as an excitation via \emph{threshold perturbation}.
This is the primary mechanism by which the optogenetic GEF perturbs Rho excitation within the region of illumination.
Second, in the absence of membrane GEF concentration changes, the Rho-GTP concentration can be raised above the threshold to trigger an excitation, for example, due to diffusive Rho-GTP transport on the membrane in the vicinity of steep Rho-GTP concentration gradients (Fig.~\ref{fig3}a,b, $t = 12\text{ min}$).
Once the Rho-GTP concentration is brought above the excitation threshold, an excitation is triggered locally.
With sufficiently strong spatial coupling, this local excitation can propagate and cause subsequent Rho excitations at distant membrane sites (Fig.~\ref{fig3}a,b, $t = 20\text{ min}$).
We refer to this as an excitation via \emph{state perturbation}.
For state perturbations, the Rho excitation threshold is primarily set by variables that are independent from light activation, specifically the endogenous GEF-Rho interaction (Supplementary Information).
In both cases, the amplitude of the Rho excitation increases with the strength of the perturbation (Extended Data Fig.~\ref{ExtFig5}c), effectively creating a \emph{threshold region} that is bound from below by the excitation threshold~\cite{McCormick.etal1991, Meron1992}.

Two strategies thus exist for propagating the Rho excitation in starfish oocytes.
On the one hand, a spatio-temporally varying threshold perturbation can guide a Rho excitation across the oocyte (\emph{guided wave}), as is the case for wild-type SCWs in meiotic occytes where the threshold perturbation is caused by a moving Ect2 front~\cite{Bischof.etal2017, Wigbers.etal2021}.
We predict that a similar guided wave can be created when the spatio-temporal guiding cue is emulated by a cell-scale optogenetic threshold perturbation pattern.
On the other hand, a prevailing steep Rho-GTP concentration gradient (initiated by, e.g., regional light activation) can give rise to state perturbations that propagate the excitation across the oocyte.
These \emph{trigger waves} are self-sustained (unguided) if the Rho-GTP concentration gradient at the wavefront is strong enough to support the superthreshold state perturbation of the medium indefinitely~\cite{Meron1992, Cross.Hohenberg1993, Murray2003} (Fig.~\ref{fig3}c).
In contrast, in the case of weak or subthreshold state perturbations (typically for a high excitation threshold or weak diffusion of Rho-GTP), the medium cannot support the trigger waves, and any kind of excitation wears off over time~\cite{Meron1992} (Fig.~\ref{fig3}d, Supplementary Information).
This predicts the formation of unguided waves whose propagation or decay depends on the strength of the initial threshold perturbation provided by local optogenetic stimuli, which then affects the later strength of state perturbations outside the stimulated regions (Fig.~\ref{fig3}c-d).

\section{Tuning oocyte response to optical stimuli}%
Our analysis indicates that threshold perturbations can be directly induced by a light stimulus, and that the response to state perturbations is affected by the endogenous excitability of the GEF-Rho system.
Thus, the outcomes of optogenetic GEF manipulations depend on the combination of threshold and state perturbations that are achievable through spatio-temporally patterned optogenetic cues.
In particular, our model analysis predicts that the following phenotypes can be produced: 
\begin{enumerate*}[label=(\roman*)]
    \item local deformations (pinches) via threshold perturbations, with diffusive fluxes (resulting in state perturbations) too weak to maintain the Rho wave;
    \item unguided SCWs (trigger waves) via threshold perturbations, where sufficiently strong state perturbations maintain the wave outside the illuminated region; and
    \item wild-type like guided SCWs after global illumination, where a gradient in the concentration of the optogenetic anchor complexes determines the threshold perturbation pattern and thus the SCW propagation axis (Extended Data Fig.~\ref{ExtFig6}).
\end{enumerate*}
In addition to these three distinct phenotypes, our model analysis shows that a transition between pinches (phenotype~(i)) and unguided SCWs (phenotype~(ii)) is possible depending on the strength of the state perturbation.
These intermediate phenotypes are characterized by a decaying Rho excitation amplitude (and thus a decaying perturbation strength due to subthreshold perturbations), giving rise to a variable propagation distance (Supplementary Fig.~S4, Supplementary Information).

\begin{figure*}[!t]
    \centering
	\includegraphics[width=1\textwidth]{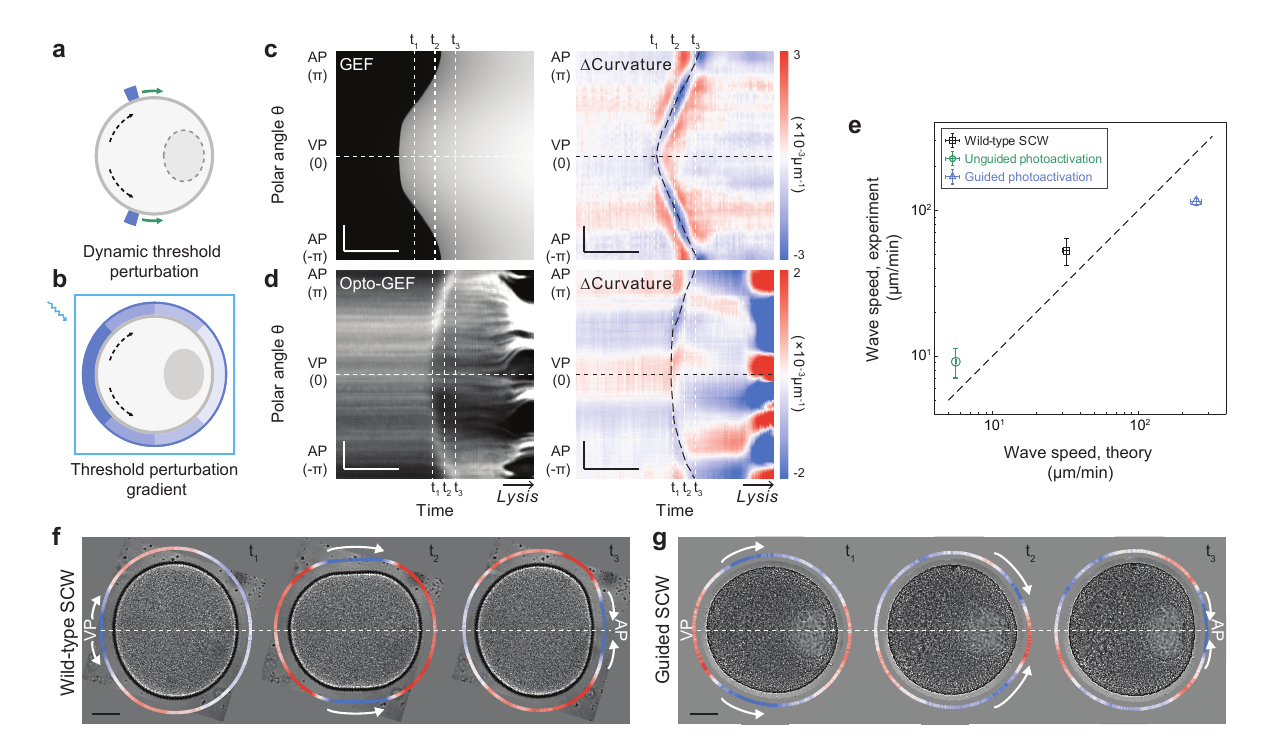}
	\caption{\textbf{Guided activation of surface contraction waves (SCWs) are experimentally realized with global illumination assays.}
        \textbf{a,}~Schematic arrangement of threshold perturbations leading to guided surface contraction waves for wild-type SCWs and 
        \textbf{b,}~for global illumination (cyan box).
        \textbf{c,}~Membrane GEF intensity (left, reconstituted from model) and curvature (right) kymographs corresponding to the experimental measurement of a wild-type oocyte undergoing SCW upon entry of meiotic cell cycle. Scale bar: $100\,\mu \text{m}$, 10~min.
        \textbf{d,}~Membrane Opto-GEF intensity (left) and curvature (right) kymographs corresponding to the experimental realization of a guided SCW through an Opto-GEF membrane gradient established by global illumination. Continuous global illumination was turned on for 45~min until the wave started. Scale bar: $100\,\mu \text{m}$, 10~min.
        \textbf{e,} Comparison of the propagation speeds of wild-type, guided and unguided SCWs. Graph shows both experimentally-measured oocyte responses and numerical predictions. Error bars represent standard deviation within each statistical group (wild-type: N=5; guided: N=4; unguided: N=5).
        \textbf{f,} Representative timelapse snapshots of the wild-type oocyte (same as (\textbf{c})). Colored rings show the instantaneous distribution of curvature along membrane. Scale bar: $50\,\mu \text{m}$. 
        \textbf{g,} Representative timelapse snapshots of the oocyte exhibiting guided SCW (same as (\textbf{d})). Colored rings show the instantaneous distribution of curvature along membrane. Scale bar: $50\,\mu \text{m}$.
        }
	\label{fig5}
\end{figure*}%

With the Ect2-based optogenetic GEF, we obtained local deformations in response to regional illumination (Fig.~\ref{fig1}e) but no unguided SCWs, even with a strong light stimulus or high Opto-GEF expression level.
This means that state perturbations in Opto-GEF-loaded oocytes are too weak to support trigger waves.
Importantly, this limit cannot be overcome by simply overexpressing active (unphosphorylated) GEF Ect2 to enhance Rho cortical excitability~\cite{Bement.etal2015, Tan.etal2020, Liu.etal2021}.
This is because overexpression, albeit lowering the excitation threshold, also leads to reduced excitation amplitudes, meaning that the ensuing state perturbations are insufficient to support trigger waves (Extended Data Fig.~\ref{ExtFig5}c, Supplementary Information).
We verified this reasoning by overexpressing active Ect2 alongside the optogenetic GEF vectors and measuring the oocyte responses to regional illumination (Methods).
Indeed, despite an enriched Rho-GTP presence at membranes, only pinching phenotype was observed in these oocytes (Extended Data Fig.~\ref{ExtFig6}a-c).

How can unguided SCWs in response to regional illumination be achieved instead?
According to our analysis, this requires a sufficiently strong state perturbation, which can be achieved by an elevated Rho-GTP diffusivity or a lowered Rho excitation threshold.
While neither of the two effects is directly controllable by optogenetic GEF, we note that in our system the endogenous GEF-Rho excitability may be indirectly influenced by Opto-GEF in the absence of light activation, for example if a small portion of the cytosolic Opto-GEF expression mediates Rho catalytic activity at membrane~\cite{Taslimi.etal2016}.
Such a ``passive effect'' (without illumination) has indeed been observed using a photosensitive GEF based on the DH domain of the Leukemia-associated RhoGEF~\cite{Jaiswal.etal2011} (LARG) in HeLa cells~\cite{Wagner.Glotzer2016,Wagner2016} (Supplementary Information).
We therefore explored a LARG-based optogenetic GEF (Opto-GEF*; Methods, Extended Data Fig.~\ref{ExtFig7}, Movie S5) to induce a stronger state perturbation compared to the Ect2-based Opto-GEF.
By modulating the light illumination and Opto-GEF* expression conditions (Supplementary Information), we indeed observed a variety of dynamic cell deformations including localized contraction (Fig.~\ref{fig4}a,c,f) and unguided SCWs that propagate across the oocytes and consistently start from the illuminated region (Fig.~\ref{fig4}b,d,g, Extended Data Fig.~\ref{ExtFig8}a-e, Movie S6).
In addition, we observed intermediate phenotypes with SCWs that only propagate over a part of the oocyte.
The propagation distance correlates with the initial threshold perturbation strength in the oocyte, as quantified via the normalized Opto-GEF* accumulation after photoactivation (Fig.~\ref{fig4}e, Extended Data Fig.~\ref{ExtFig8}f,g, Supplementary Information).
We conclude that using Opto-GEF* indirectly enhances the effect of the state perturbations, possibly through effective changes to the autocatalytic exchange rate~\cite{Wagner2016} through our model analysis (Supplementary Information).

\begin{figure*}[!t]
\centering
	\includegraphics[width=1\textwidth]{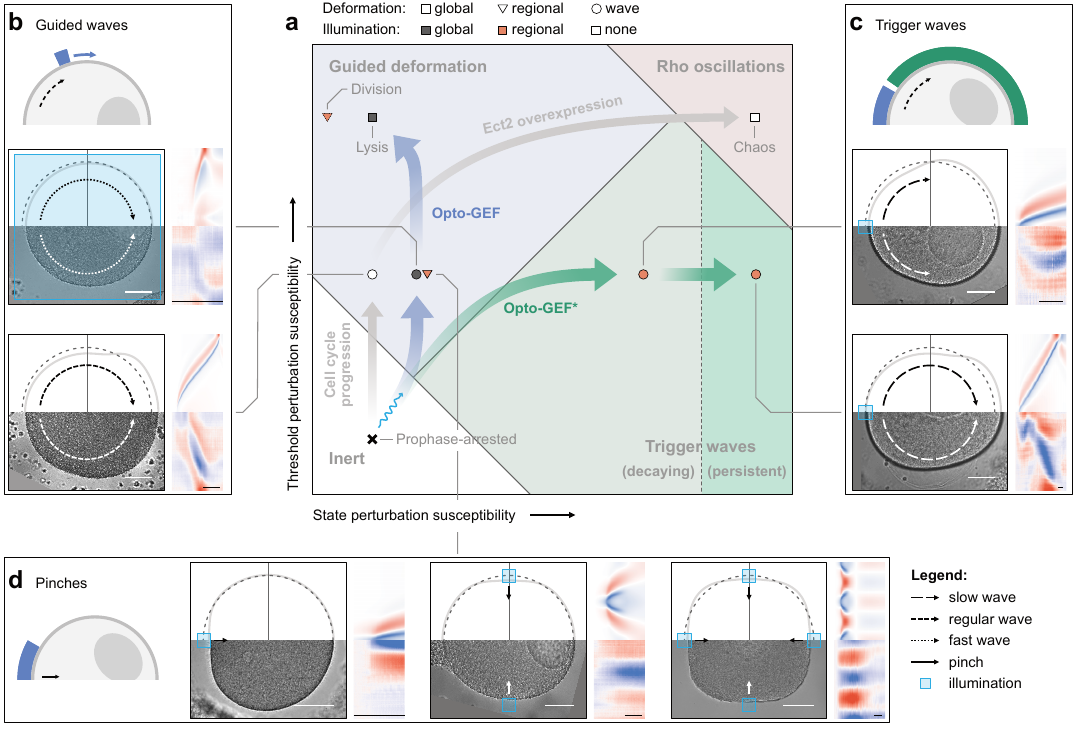}
	\caption{\textbf{Qualitative map of oocyte shape deformations.}
        \textbf{a,}~Different deformation phenotypes are obtained for varying threshold and state perturbation susceptibility. Unmodified oocytes before the nucleotide envelope breakdown are inert (white region) and do not deform at all. Increasing the threshold perturbation susceptibility (experimental realization by Opto-GEF addition, blue shaded arrow) enables generating guided waves and pinch deformations (blue region) after photoactivation (represented by a wavy blue arrow). Increasing the state perturbation susceptibility (experimental realization by Opto-GEF* addition, green shaded arrow) enables unguided waves of varying propagation distance (green region). Chaotic and spiraling dynamics (gray) are obtained when both threshold and state perturbation susceptibility are high, as observed for Ect2 overexpression~\cite{Wigbers.etal2021}.
        \textbf{b,}~Guided threshold perturbations cause traveling waves, where guidance can be either due to an optogenetic guiding cue (membrane anchor concentration, top) or due to a native cue (traveling Ect2 front in WT oocytes, bottom).
        \textbf{c,}~For strong state perturbation, regional illumination results in unguided traveling waves. The propagation distance of the unguided wave depends on the state perturbation susceptibility.
        \textbf{d,}~For weak or subthreshold state perturbations, regional illumination results in localized pinches. Different cell shapes can be produced by changing the spatial pattern of the light stimuli.
        For each snapshot and curvature kymograph, the top half is obtained from simulations, while the bottom half is an experimental realization. Scale bars: 10~min (kymographs), $50\,\mu\text{m}$ (snapshots).
	}
	\label{fig6}
\end{figure*}%

Lastly, from the model follows that a SCW can be created using light stimuli without unguided propagation of the excitation (Supplementary Information).
Here, the wave needs to be guided by spatio-temporally varying threshold perturbations.
In optogenetically treated oocytes, we verified that the guiding cue can be provided by the Opto-GEF concentration front that forms in oocytes with an anchor-complex concentration gradient along the AP-VP axis poised under global illumination (Extended Data Fig.~\ref{ExtFig6}d,e, Supplementary Information).
The resulting SCW robustly travels from VP to AP (Fig.~\ref{fig5}, Movie S8).
Similar to the wild-type meiotic SCW that is guided by a decaying gradient of Cdk1~\cite{Bischof.etal2017}, a large-scale cytoplasmic flow that is symmetric across the AP-VP axis~\cite{Klughammer.etal2018} is created for the Opto-GEF-induced guided SCW (Methods, Extended Data Fig.~\ref{ExtFig9}a,b).

Taken together, the three different types of SCWs -- unguided waves after regional illumination, guided waves after global illumination, and wild-type guided waves -- are each caused by a different mechanism and therefore travel at vastly different speeds (Fig.~\ref{fig5}e).
For guided waves, the wave speed is determined by the guiding cue (anchor gradient slope for light-induced guided SCWs and Ect2 front speed for wild-type SCWs~\cite{Wigbers.etal2021}).
In contrast, the speed of unguided waves is predominantly determined by the net mass redistribution of Rho-GTP in the vicinity of activity peaks.
In particular, the diffusive flux of Rho-GTP is counteracted by in-plane advective flows focusing on regions of high contractility which, by virtue of Rho-induced actomyosin activity~\cite{Narumiya.etal2009}, coincide with the Rho activity peaks (Supplementary Information).
We emphasize that at the heart of this tunability of cell shape dynamics lies the versatility of the Rho protein system that allows both threshold perturbations and state perturbations, which can be controlled through an optogenetic GEF and predicted by the chemo-mechanical model analysis.

\section{A map of deformation phenotypes}%
Our framework suggests that all observed deformation phenotypes can be interpreted as a combination of threshold and state perturbations.
The susceptibility of a cell to either mechanism generically depends on the endogenous GEF-Rho system properties.
The expression of optogenetic GEFs indirectly enhances the state perturbation susceptibility of a cell.
Light illumination induces spatio-temporal threshold perturbations, and the resulting deformation phenotype depends on the cell's susceptibility to both perturbation mechanisms.
In Fig.~\ref{fig6}a, we present a qualitative map based on the susceptibilities to threshold and state perturbations and assigned the different deformation phenotypes as obtained from numerical simulations (Supplementary Information).
Here, the minimal perturbation required to cause an excitation decreases as the corresponding perturbation susceptibility increases.

The threshold and state perturbation susceptibilities are mesoscopic quantities determined by a combination of experimental and biological conditions and as such they are difficult to quantify.
Therefore, the map is meant to provide essential qualitative insights for the experimental designs.
For example, the prediction that a stronger susceptibility is needed for trigger waves motivated our use of a different optogenetic GEF species (Opto-GEF*).
Another example is the observation of occasional oocyte lysis events following global illumination of oocytes with Opto-GEF or Opto-GEF* assay expression (Extended Data Fig.~\ref{ExtFig9}d, Extended Data Fig.~\ref{ExtFig10}, Movies S9-S11).
These extreme actomyosin contractions that led to an abrupt oocyte lysis are a result of enhanced Rho activity bursts, consistent with our prediction of an elevated threshold perturbation susceptibility for oocytes with high expression levels of photosensitive tag complexes (Supplementary Information).
In Fig.~\ref{fig6}b-d, we list examples of experimentally observed phenotypes side-by-side with quantitatively matching results from numerical simulations.
In simulations, we realized changes of threshold and state perturbation susceptibilities by varying the Opto-GEF concentration and the autocatalytic nucleotide exchange rate, respectively.
Note that the latter is not a unique choice (Supplementary Information); however, as an effective coarse-grained parameter, it encompasses the relevant nonlinear and indirect Rho-GEF interactions.

In summary, optogenetic control of Rho activity has been realized in a variety of model organisms~\cite{Wagner.Glotzer2016, Kamps.etal2020, Cavanaugh.etal2020,Rich.etal2020, Herrera-Perez.etal2023, Seze.etal2023} and is a powerful tool for gaining external control over cell shape dynamics.
In the past, the large variety of control parameters associated with this tool (e.g., expression level, illumination condition, and use of exogenous constructs) has posed a general difficulty for quantifying and predicting optogenetically induced effects systematically and across different systems.
Here, we have shown that optogenetic control is a highly versatile and predictable method for altering the shape of cells with high spatial and temporal precision.
We introduced an optogenetic guanine nucleotide exchange factor (Opto-GEF) that modulates the Rho activity on the membrane, allowing us to induce localized actomyosin contractility with light stimuli.
Using a theoretical model for the protein dynamics and the chemo-mechanical coupling, we identified two distinct mechanisms via which a light stimulus can induce cell deformations  -- threshold perturbations and state perturbations, each leading to Rho excitation.
We confirmed both mechanisms experimentally using different variants of the optogenetic GEF and demonstrated the versatility of our optogenetic method by generating predicted custom oocyte shape dynamics with light.
Finally, we provided a qualitative map that is based on the two perturbation susceptibilities to bring the various optogenetically or meiotically activated phenotypes together into one unified framework.

Although our theoretical model entails a wide range of processes relevant to the optogenetic control of the oocyte's shape dynamics (reaction--diffusion dynamics of optogenetic and endogenous proteins, cortical flows, and shape changes), we emphasize that it is by design an effective description of the actual dynamics that relies on a number of assumptions and simplifications.
Most importantly, we disregard realistic actin dynamics and cytoplasmic flows, which could affect the reaction--diffusion dynamics resulting in the Rho excitation.
In addition, since the molecular origin of the nonlinear contribution to the Rho dynamics has not yet been fully uncovered~\cite{Chen.etal2020}, the corresponding model term (autocatalytic Rho activation) is based on mesoscopic observations and previous studies~\cite{Bement.etal2015, Goryachev.etal2016, Wigbers.etal2021, Michaud.etal2022}.
While mismatches between the model description and the actual dynamics may affect the quantitative results, we expect our qualitative analysis to be independent from the specific molecular origin of the nonlinear dynamics, since similar mesoscopic dynamics can often be realized by multiple different and possibly redundant microscopic interactions in biological systems~\cite{Goryachev.Leda2017, Brauns.etal2023}.

From an experimental perspective, classifying deformation phenotypes as shown in Fig.~\ref{fig6}a illustrates how more complex oocyte shapes can be induced.
To ensure a localized response to light stimuli, the threshold perturbation needs to be enhanced, whereas the state perturbation susceptibility should be attenuated.
Interestingly, highly localized low-amplitude Rho oscillations are also observed prior to meiotic oocyte division in a narrow band around the cell equator~\cite{Su.etal2014,Bement.etal2015}.
This localized signaling might be achieved by an intracellular regulatory mechanism that modulates the perturbation susceptibilities in space and time.
In general, our results suggest that extreme cell deformations can be generated by optimizing the optical stimuli to induce persistent and localized deformations, and a permanent restructuring of the cell cortex~\cite{Sedzinski.etal2011}.
More broadly, our optogenetic method of applying threshold and state perturbations may be employed to both control cortex contractility and guide curvature-generating proteins in synthetic cells as a means to develop cell deformation, motility, and division in synthetic biology~\cite{Toettcher.etal2011, Schwille2019, Baldauf.etal2022}.





\section{Acknowledgements}%
This research was supported by Sloan Foundation Grant (G-2021-16758) and a National Science Foundation CAREER Award (Grant No. PHYS-1848247) to N.F..
E.F.~acknowledges financial support by the German Research Foundation (DFG) through SFB1032 (Project ID No. 201269156), and support from Germany’s Excellence Strategy, Excellence Cluster ORIGINS, EXC-2094-390783311.
J.L.~was supported by the MathWorks Science Fellowship and the ELBE Fellowship.
T.B.~acknowledges support from the Joachim Herz Foundation.
J.R.~was supported by the post-course research funding from the 2022 MBL Physiology course.
We thank Peter~J.~Foster and Ching Yee Leung for helpful discussions, Pu Zheng for the advice and help on T7 in vitro transcription, Charles Brad Shuster for valuable discussions of GEF inhibition tools, and the 2022 MBL Physiology course in Woods Hole for valuable inputs.

\section{Author contributions}%
N.F., E.F., J.L.~and T.B.~conceived and designed the study.
J.L.~and T.H.T.~designed the experimental assays.
J.L.~and J.R.~performed experiments with inputs from Y.C., T.H.T.~and S.Z.S..
J.L.~analysed the data.
T.B.~and A.Z.~conceived the theoretical analysis.
T.B.~performed the numerical simulations.
J.L., T.B., A.Z., E.F.~and N.F.~wrote the manuscript with inputs from all authors.

\section{Competing interests}%
The authors declare no competing interests.

\clearpage
\bibliography{bibliography}

\clearpage

\section{Methods}
\subsection{Starfish oocytes handling}%
The bat stars (\textit{Patiria miniata}) we used were purchased from South Coast Bio-Marine or Marinus Scientific. 
The animals were procured from Southern California coast and kept in 15\degree C seawater tanks. 
To collect oocytes, a small incision was made in a female starfish with a scalpel. 
The ovary gonads were then extracted with a pair of tweezers and cut up carefully with scissors to release individual oocytes. 
To ensure that oocytes are properly arrested at prophase, we collect them through sedimentation after washing twice with calcium-free seawater and kept at 15\degree C in regular seawater until usage. 

For the optogenetic experiments, we used oocytes arrested in prophase. 
For acquisition of oocytes entering meiosis, we induced maturation by applying 1-methyl adenine (1MA, Acros Organics) at $10\,\mu\text{M}$. 
All of the experimental results were obtained from multiple collections where oocytes were used within 48~h from the time of collection. 

\subsection{Optogenetic constructs and imaging assays}%
Our design of the optogenetic switch was based on the reversible dimerization between the light-sensitive cryptochrome2 protein, Cry2PHR, and its partner protein, CIBN, that were reported from earlier work~\cite{Kennedy.etal2010, Taslimi.etal2016}. 
Both constructs were gifts from Chandra Tucker (CRY2PHR(W349R)-mCherry: Addgene plasmid \#75370; http://n2t.net/addgene:75370; RRID: Addgene\_75370. pCIBN(deltaNLS)-pmGFP: Addgene plasmid \#26867; http://n2t.net/addgene:26867; RRID: Addgene\_26867). 
In the gifted sequences, the CIBN domain was fused to a CaaX polybasic sequence, allowing for its localization to plasma membranes and the recruitment of CRY2PHR domains from the cytosol. 
To induce changes of GEF catalytic activities with this switch, we made use of the previously identified functional domains of the Ect2 GEF~\cite{Su.etal2011, Su.etal2014} and the exogenous Larg GEF~\cite{Wagner.Glotzer2016}, fusing the corresponding catalytic GEF DH sequences to Cry2PHR.
The full sequence of this Ect2 (with a point mutation outside the DH sequence compared to the endogenous Ect2) was also used in the control implementation of Opto-GEF assay with an elevated cortical excitability.
The two GEF constructs were gifts from G.~von Dassow (pCS2+mCh-SpEct2 T808A mini \#1) and Michael Glotzer (2XPDZ\_mCherry\_LARG DH (aa 766-997): Addgene plasmid \#80407; http://n2t.net/addgene:80407; RRID:Addgene\_80407).
All sequences were next cloned into pCS2+8 vectors to enable the synthesis of mRNAs using the mMessage mMachine SP6 Transcription Kit (Invitrogen) and the Poly(A) Tailing Kit (Invitrogen) after DNA linearization. 

Three photoactivation assays were created in our experiments, by expressing mixed Cry2PHR and CIBN mRNAs in oocytes: the eGFP-tagged CIBN membrane anchor paired with
\begin{enumerate*}[label=(\roman*)]
    \item mCherry-tagged control Cry2PHR (Opto-Null),
    \item Ect2 DH-fused Cry2PHR (Opto-GEF),
    \item and Larg DH-fused Cry2PHR (Opto-GEF*).
\end{enumerate*}
In order to enable the imaging of Rho activity with a GFP-tagged rGBD reporter~\cite{Benink.Bement2005}, we performed a separate control experiment where the CIBN-CaaX sequence was cloned into a fluorescence-untagged construct.
The mRNA of the untagged CIBN protein was then mixed with mRNAs of the GFP-rGBD (a gift from William Bement, Addgene plasmid \#26732; http://n2t.net/addgene:26732; RRID:Addgene\_26732) and mCherry-Cry2PHR constructs for expression.
All mRNA mixtures were first made with approximately equal concentrations from each component, then the ratio was adjusted according to oocyte batch-specific responses (see Supplementary Information for details).
Using custom-made chambers and air-filled glass needles, the mRNA mixtures were microinjected into individual oocytes and incubated overnight (15-24~h) to ensure strong fluorescence signals for the subsequent imaging and photo-manipulation~\cite{Dassow.etal2018}.

\subsection{Myosin imaging and functional inhibition assay}%
We performed myosin imaging by expressing the heavy chain protein of non-muscle myosin II (MHC-GFP), a gift construct from Peter Lenart.
The MHC-GFP mRNA was synthesized using the mMessage mMachine T7 Transcription Kit (Invitrogen) and the Poly(A) Tailing Kit (Invitrogen) after DNA linearization.
The functional inhibition was performed by pre-incubating prophase-arrested starfish oocytes in seawater with 200~$\mu$\text{M} (-)blebbistatin (Sigma-Aldrich).
The blebbistatin was dissolved in Dimethyl sulfoxide (DMSO, Sigma-Aldrich) before being suspended in the seawater. 
The final DMSO concentration is 0.4\% both for blebbistatin treatment and for control incubation. 
The oocytes were incubated for 2-4~h before imaging experiments.
For MHC-GFP signals, the fluorescent images were first segmented in ilastik~\cite{Berg.etal2019} to remove bright clusters, then processed for near-membrane intensities.

\subsection{Photoactivation with confocal microscopy}%
We performed the photoactivation experiments primarily using the two-color imaging with 488~nm and 561~nm laser excitations on a Zeiss LSM 710 laser scanning confocal microscope (40X water objective, NA 1.1 corr).
For a small number of the experiments where strict laser power controls were not essential, we also acquired results from a Zeiss LSM 700 laser scanning confocal microscope (40X water objective, NA 1.3, 491~nm and 555~nm laser excitation).
Both microscope rooms were maintained at 22-25\degree C.
The laser scanning confocal microscopy was most efficient in creating regional optical excitations within a single z-plane during the image acquisition. 
Therefore, we custom-made imaging chambers from double-sided tapes (3M) and coverslips that have a fixed height ($\approx 100\, \mu \text{m}$). 
When held by the chamber, the oocytes were slightly compressed on the top and bottom (Extended Data Figure~\ref{ExtFig4}a) and prevented from positional drifts, while the bulk in between remained flexible.
We then performed illumination and acquisition on a single z-plane near the center of the bulk where the shape response of the oocyte after photoactivation was most visible.

Our photoactivation pipeline works as following. 
After loading the incubated oocytes into the imaging chamber, a first selection of oocytes showing strong fluorescence expression in the Cry2PHR channel was made. 
For \textit{local} activation, we avoided direct imaging with the 488~nm laser and set up a ``photo-bleaching'' function in the acquisition software (Zeiss Zen Black 2012) to scan specified regions of interest using the 488~nm laser. 
The scan was set to the same frequency as the time series acquisition in the Cry2PHR fluorescence channel ($1\,\text{s}^{-1}$).
For \textit{global} activation, we set up the photoactivation using the dual-channel imaging with 488~nm light and the Cry2PHR fluorescence channel (561~nm).
To cope with photobleaching at long times in the case of \textit{global} activation, we imaged at a lower fixed frequency ($0.1\,\text{s}^{-1}$). 
For both acquisition settings, other parameters such as the pixel dwell time, scanning laser power and detector conditions were set to be the exact same.
In our experiments we found no traces of qualitative differences caused by the fast and slow activation and acquisition frequency.

\subsection{Analysis of photoactivation data}%
This section describes the processing of time-lapse image data collected from imaging and photoactivation experiments. 
Image formatting was done with the Fiji software and analyses were performed using customized MATLAB (version R2018b) scripts.
We processed the numerical data (see next sections) using the same pipeline with minor modifications.

\paragraph{Generation of intensity and curvature kymographs.}
To generate a kymograph that describes signal propagation along the cell boundary, the contour of the oocyte was first identified using the “bwboundaries” function in MATLAB. 
The pixel-based intensity kymograph was then determined as a moving box on the boundary that averages the intensity of all pixels inside the box. 
For datasets with low signal-to-noise ratio, we average instead from pixels that have top 1/3 brightness in the box to enhance kymograph contrast.
To obtain the curvature kymograph, pixels of a moving segment of the boundary were smoothed and interpolated (“interp1” function in MATLAB) with a five-point spline curve. 
The second-order spatial derivative of the center of the spline was then retrieved as the local curvature value, 
\begin{linenomath}
\begin{equation*}
    C(s) = \sqrt{\left(\frac{\mathrm d^{2}x}{\mathrm ds^{2}}\right)^{2}+\left(\frac{\mathrm d^{2}y}{\mathrm ds^{2}}\right)^{2}}.
\end{equation*}
\end{linenomath}
An additional fitting step (“interp1” function in MATLAB) combining the boundary pixel positions was done to map the obtained kymographs from pixel-based cartesian coordinates to angle-based polar coordinates.

\paragraph{Quantification of compartmentalized intensity kinetics.}
To evaluate the recruitment of Cry2PHR photosensitive tag, we segmented the oocyte image into cytosol and membrane compartments in the analyses. 
The membrane compartment was generated by first extracting the oocyte contour and then rescaling the contour shape to get a collection of pixels proximal to the cell surface.
The cytosol compartment contains rest of the pixels that are not proximal to the cell surface but inside the oocyte contour. 
The nucleus region was excluded from the cytosol compartment through a second segmentation.
In \textit{global} activation, averages of pixels inside each compartment were taken to extract the spatially-averaged mean intensity kinetics (Extended Data Figure~\ref{ExtFig3}).
In the case of \textit{local} activation within a specified region of interest (ROI), the compartments were combined with the location of the ROI to generate the pixel collections that were then used to extract the intensity kinetics (Extended Data Figure~\ref{ExtFig2}). 

In the cases when membrane recruitment kinetics are to be quantified (e.g. \textit{local} membrane recruitment, Fig.~\ref{fig1}d) or compared (e.g. the continuous and pulse \textit{global} activation scenarios, Fig.~\ref{fig2}b; the continuous \textit{global} activation of two biochemical species, Fig.~\ref{fig2}d), we extract the fractional membrane and cytosol responses using the nucleus-excluded compartmentalized intensities. 
In these cases, the fractional responses were normalized to the maximal values within the data group. 
The baseline of each fractional time series was defined as the average value before light activation was applied, and was set to 0 for membrane fraction (1 for cytosol fraction).

In the case where initial protein expressions are to be carefully evaluated (e.g. comparison of \textit{local} membrane recruitment kinetics across Opto-GEF* assays, Extended Data Figure~\ref{ExtFig8}), we take the membrane mean intensity time series without fractional normalization with cytosolic readings.
In this case, the membrane intensities were normalized to the pre-activation baseline value, so to start at 1 when light was applied.
The normalized maximums of the membrane intensity in these time series were then taken as fold recruitment values that are specific to the expression and activation profiles of individual oocytes (Fig.~\ref{fig4}e, Extended Data Figure~\ref{ExtFig8}g).

\paragraph{Quantification of edge contraction kinetics.}
In the case of \textit{local} activation near oocyte membranes, we used the edge displacements to evaluate and compare timescales of the mechanical responses after photoactivation.
As before, the baseline of the contraction time series was defined as the contour position before activation was applied. 
The concurrent displacement were then defined as the point-to-point distance between the current and the baseline contours, marking the centre of \textit{local} location on the boundary (Fig.~\ref{fig2}c, Extended Data Figure~\ref{ExtFig3}f, Extended Data Figure~\ref{ExtFig7}a-b).
In the cases when edge contraction kinetics are to be quantified, the edge displacement time series was normalized to the maximal displacement that took place in the time series (Fig.~\ref{fig2}c, Extended Data Figure~\ref{ExtFig3}f).

\paragraph{Quantification of wave speeds.}
We used the cell surface curvature time series to extract key SCW quantities such as propagation distances (Fig.~\ref{fig4}e) and mean propagation speed (Fig.~\ref{fig5}e).
To obtain the SCW propagation distance, for each time point, an inverse gaussian peak was fitted to the regional curvature profile corresponding to the contractile front.
The distance between the two inverse peak positions fitted from both ends of the contractile front was taken as the concurrent SCW propagation distance.
The maximal propagation distance in time was then extracted as the final propagation distance that is specific to an individual Opto-GEF* oocyte under \textit{local} activation.

To calculate the SCW speed, we located the curvature minimum for each polar angle boundary position on the curvature kymograph.
A linear fit between the boundary position and temporal appearance was then performed for the time segment when a contractile front visibly propagates. 
The mean wave speed was then averaged from the fit results for all individual oocytes across different SCW scenarios (wild-type, guided and unguided photoactivation).

\paragraph{Fitting of photoactivation parameters}
For measuring the “on” and “off” half-times for the optogenetic switch, the normalized fractional kinetics were fitted using exponentials $I(t) = 1-e^{-t \cdot \ln{2}/t_\text{on}}$, $I(t) =I(t_{\infty})-(I(t_{\infty})-I(0)) \, e^{-t\cdot\ln{2}/t_\text{off}}$ and the “lsqcurvefit” function in MATLAB.
The ratio of the membrane attachment and detachment rates of the optogenetic tag complexes can be quantified from the Opto-GEF binding kinetics after regional pulse~(light on for $1\,\text{s}$) illumination (Fig.~\ref{fig2}c).
The photoactivation and cytosolic diffusion rates, on the other hand, can be estimated from global illumination (entire oocyte) experiments, as the time scale of membrane recruitment is limited by diffusion of cytosolic complexes towards the membrane at long times (Fig.~\ref{fig2}b, Extended Data Fig.~\ref{ExtFig3}a-c). 
Similar dynamics are observed when using the fully functional variant of the optogenetic tags (Opto-GEF), where the cortex contraction can be tracked simultaneously (Fig.~\ref{fig2}c, Supplementary Fig.~S7e-f).

\paragraph{Decomposition of oocyte shapes with 2D Fourier polar modes.}
We used the 2D Fourier modes as a common method~\cite{Fregin.etal2019, Perez-gonzalez.etal2019} for evaluating and dissecting shape dynamics after perturbation.
In polar coordinates, the contour line as an angle-dependent radius function $r(\theta)$ can be decomposed into its (spatial) frequency components: 
\begin{linenomath}
\begin{equation*}
r(\theta) = a_{0} + \sum_{k=1}^{\infty} \bigl(a_{k} \cos(k\,\theta) + b_{k} \sin(k\,\theta)\bigr)
\end{equation*}
\end{linenomath}
These trigonometric functions are shape eigenmodes with symmetry axis parallel ($\cos(k\,\theta)$) or diagonal ($\sin(k\,\theta)$) with respect to a reference axis. 
To obtain the mode amplitudes $a_{k}$ and $b_{k}$ from experimental data, we treated the oocyte contour as a closed $N$-point polygon and extracted the center of mass $x_{c}, y_{c}$ ($N$ equals the number of contour pixels.
Points on the polygon contour $x_{n}{-}x_c$, $y_{n}{-}y_c$ were then transformed to polar coordinates $r_{n}, \theta_{n}$, and a discrete Fourier transform was performed to retrieve the Fourier coefficients: 
\begin{linenomath}
\begin{align*}
a_{k} &= \frac{1}{\pi} \sum_{n=0}^{N-1} r_{n} \, \cos(k\,\theta_{n}) \, \delta\theta_{n} & \text{for} \quad k &\geq 0 \,, \\
b_{k} &= \frac{1}{\pi} \sum_{n=0}^{N-1} r_{n} \, \sin(k\,\theta_{n}) \,\delta\theta_{n} & \text{for} \quad k & \geq 1 \, .
\end{align*}
\end{linenomath}
$\delta\theta_{n}$ was taken as a step difference $\frac{1}{2}(\theta_{n}-\theta_{n-1}) + \frac{1}{2}(\theta_{n+1}-\theta_{n})$.
The accuracy of the algorithm was verified by reconstructing the original oocyte shape from the decomposed mode amplitudes.
Since the coefficients $b_{k}$ quantifies “off-axis” dynamics, we chose the reference axis as the one that minimizes the “off-axis” coefficients $b_{k}$ and focused on the resulting $a_{k}$ dynamics.
For wild-type SCWs, this minimization approach nicely recovered the VP-AP axis. 
For a detailed discussion of the results obtained from the Fourier shape decomposition, we refer to Supplementary Information.

\paragraph{Quantification of mean deformation rates.}
For the myosin inhibition assay, the oocyte deformation rate was calculated for quantifying the myosin inhibition effect for meiotic surface contraction waves and for quantifying the local Opto-Ect2/myosin activation effect for photoactivatable oocytes.
We used the definition of oocyte deformation rate from previous work~\cite{Foster.etal2022} as a reference.
At each time point, the deformation rate at a membrane site is calculated by tracking the magnitude of decrease of the normalized radius between this site and the center of mass for this oocyte.
The radius at each site is normalized by the site's distance from the center of mass (radius) before the onset of contractions.
For meiotic surface contraction waves, the peak deformation rate is calculated by taking the maximum value as the wave passes (the maximum value for all membrane sites throughout all time points) and evaluated both for oocytes without blebbistatin inhibition (DMSO treatment) and oocytes with 200~$\mu\text{M}$ blebbistatin inhibition.
For the local light illumination (488~nm), the mean deformation rate is measured from all membrane sites inside the illumination range (averaged for all time points after illumination).
Noise floor (gray) is defined by the fluctuation errors in traced oocyte outlines and is segmentation-dependent.

\paragraph{Quantification and visualization of cytoplasmic flows.}
We used PIV (particle image velocimetry) analysis to extract cytoplasmic flows from the time-lapse bright field images.
The bright field images were corrected from lab frame to the center-of-mass frame of reference by first extracting the time-lapse center-of-mass trajectory from the raw images.
Cytoplasmic flow velocity fields were generated from the corrected time-lapse images using the PIVlab software in MATLAB.
Streamline visualization of the cytoplasmic velocity field was realized using the Python software package \texttt{streamplot} available in the matplotlib library~\cite{Hunter2007}.

\subsection{Photoactivation modeling}%
In this section, we specify the model equations and parameters for the chemical modules. For a detailed description of the model, we
refer to Supplementary Information. The modules describe the various interactions of cytosolic and membrane-bound species and account for their mass-conservation as an important property of many biological systems \cite{Halatek.Frey2018,Brauns.etal2020}. For simplicity, we focus on modeling the chemo-mechanical dynamics of the sea star oocytes in two spatial dimensions. Therefore, the considered bulk area corresponds to a cross-section of the oocyte volume at the experimentally observed focal plane. It is bounded by an enclosing curve representing the membrane.

\paragraph{Reaction-diffusion dynamics for the wild-type Ect2 module.}
To model the native Ect2 dynamics, and in particular the coupling to the upstream Cdk1 guidance, we extend a model previously proposed to study SCWs in sea star oocytes~\cite{Wigbers.etal2021}.
In this model, we distinguish between a phosphorylated ($u_\text{Ep}$) and an unphosphorylated state ($u_\text{E}$) of Ect2 in the cytosol, and a membrane-bound state ($u_\text{e}$).
The corresponding reaction-diffusion equations are
\begin{linenomath}
\begin{align*}
    \partial_t u_\text{Ep} &= D_\text E \nabla^2 u_\text{Ep} + f_\text{Ep}(u_\text{Ep}, u_\text{E}), \\
    \partial_t u_\text{E} &= D_\text E \nabla^2 u_\text{E} + f_\text{E}(u_\text{Ep}, u_\text{E}), \\
    \partial_t u_\text{e} &= D_\text e \nabla_\mathcal{S}^2 u_\text{e} - \nabla \cdot \bigl(\mathbf v \, u_\text{e}\bigr)+ \\
    &\qquad \qquad +f_\text{e}\bigl({\left.u_\text{Ep}\right|_\mathcal{S}}, {\left.u_\text{E}\right|_\mathcal{S}}, u_\text{e}\bigr),
\end{align*}
\end{linenomath}
where $D_\text E$ and $D_\text e$ denote the Ect2 diffusion constants in the cytosol and on the membrane, respectively, and $\nabla_\mathcal{S}$ is a spatial derivative along the membrane $\mathcal{S}$.
The advection term $\nabla \cdot (\mathbf v \, u_\text{e})$ accounts for in-plane advection of proteins due to cortical contractions that hydrodynamically couple to the membrane~\cite{Nishikawa.etal2017} where the advection velocity $\mathbf v$ is determined from an active stress relation (Supplementary Information).
The phosphorylation kinetics are assumed to follow Michaelis-Menten kinetics with Michaelis-Menten constant $K_\text{p}$.
The Cdk1 concentration dependence is accounted for by an effective phosphorylation rate $k_\text{Cdk1}$.
Dephosphorylation is assumed to happen spontaneously ($k_\text{dP}$) and via a feedback involving the unphosphorylated Ect2 concentration ($k_\text{fb}$).
Unphosphorylated Ect2 obeys simple linear binding and unbinding to the membrane, with rates $k_\text{on,e}$ and $k_\text{off,e}$, respectively.
The corresponding reaction terms including the (de-)phosphorylation and membrane binding of Ect2 are
\begin{linenomath}
\begin{align*}
    f_\text{Ep}(u_\text{Ep}, u_\text{E}) &= \frac{k_\text{Cdk1} \, u_\text{E}}{K_\text{p} + u_\text{E}} + \\ & \qquad - (k_\text{dP} + k_\text{fb} u_\text{E}) \cdot u_\text{Ep} \, , \\
    f_\text{E}(u_\text{Ep}, u_\text{E}) &= -\frac{k_\text{Cdk1} \, u_\text{E}}{K_\text{p} + u_\text{E}} + \\ & \qquad + (k_\text{dP} + k_\text{fb} u_\text{E}) \cdot u_\text{Ep} \, , \\
    f_\text{e}\bigl({\left.u_\text{Ep}\right|_\mathcal{S}}, {\left.u_\text{E}\right|_\mathcal{S}}, u_\text{e}\bigr) &= k_\text{on,e} \, {\left.u_\text{E}\right|_\mathcal{S}} - k_\text{off,E} \, u_\text{e} \, .
\end{align*}
\end{linenomath}
The cytosolic dynamics are coupled to the membrane dynamics via Robin boundary conditions
\begin{linenomath}
\begin{align*}
    D_\text{Ep} \, \hat{\mathbf{n}}_\mathcal{S} \cdot \nabla {\left.u_\text{Ep}\right|_\mathcal{S}} &= 0 \, , \\
    D_\text{E} \, \hat{\mathbf{n}}_\mathcal{S} \cdot \nabla {\left.u_\text{E}\right|_\mathcal{S}} &= - k_\text{on,e} \, {\left.u_\text{E}\right|_\mathcal{S}} + k_\text{off,E} \, u_\text{e} \, ,
\end{align*}
\end{linenomath}
where the vector $\hat{\mathbf{n}}_\mathcal{S}$ is the normal vector at the surface.
These boundary conditions implement the reactive coupling between bulk and boundary species and ensure mass conservation of Ect2 in the system. 

\paragraph{Reaction-diffusion dynamics for the optogenetic GEF module.}
In addition to the native Ect2, the Rho activation can be mediated by optogenetic GEF in the photosensitive oocytes.
Here, we distinguish between a photoactivated state ($u_\text{Ga}$) and an inactive state ($u_\text{G}$) of the optogenetic GEF in the cytosol, and a membrane-bound state ($u_\text{g}$).
To describe the reaction-diffusion dynamics, we use the following model:
\begin{linenomath}
\begin{align*}
    \partial_t u_\text{Ga} &= D_\text G \nabla^2 u_\text{Ga} + f_\text{Ga}(u_\text{Ga}, u_\text{G}), \\
    \partial_t u_\text{G} &= D_\text G \nabla^2 u_\text{G} + f_\text{G}(u_\text{Ga}, u_\text{G}), \\
    \partial_t u_\text{g} &= D_\text g \nabla_\mathcal{S}^2 u_\text{g} - \nabla \cdot \bigl(\mathbf v \, u_\text{g}\bigr) + \\ & \qquad + f_\text{g}\bigl({\left.u_\text{Ga}\right|_\mathcal{S}}, {\left.u_\text{G}\right|_\mathcal{S}}, u_\text{g}\bigr).
\end{align*}
\end{linenomath}
Here, $D_\text G$ and $D_\text g$ denote the GEF diffusion constants in the cytosol and on the membrane, respectively.
Based on experimental evidence, we assume that only the photoactivated state can bind to the membrane at a rate $k_\text{on,g}$, and binding requires the presence of membrane anchors ($u_\text{a}$).
Inactive GEF is thought to be activated by illumination in the cytosol.
To emulate this process, we assume the GEF activation rate $k_\text{a}$ to be modulated by a field $I(\mathbf{x}, t)$ that represents the light intensity.
Both membrane binding and light-induced activation are reversible processes, with rates $k_\text{off,g}$ and $k_\text{d}$ for unbinding and deactivation, respectively.
Altogether, the mass-conserving reaction terms read
\begin{linenomath}
\begin{align*}
  f_\text{Ga}(u_\text{Ga}, u_\text{G}) &= I \cdot k_\text{a} \, u_\text{G} - k_\text{d} \cdot u_\text{Ga} \, ,\\
  f_\text{G}(u_\text{Ga}, u_\text{G}) &= -I \cdot k_\text{a} \, u_\text{G} + k_\text{d} \cdot u_\text{Ga} \, ,\\
 f_\text{g}\bigl({\left.u_\text{Ga}\right|_\mathcal{S}}, {\left.u_\text{G}\right|_\mathcal{S}}, u_\text{g}\bigr) &= u_\text{a} \cdot k_\text{on,g} \, {\left.u_\text{Ga}\right|_\mathcal{S}} + \\ & \qquad - k_\text{off,g} \, u_\text{g} \, ;
\end{align*}
\end{linenomath}
supplemented with boundary conditions
\begin{linenomath}
\begin{align*}
D_\text{Ga} \, \hat{\mathbf{n}}_\mathcal{S} \cdot \nabla {\left.u_\text{Ga}\right|_\mathcal{S}} &= - u_\text{a} \cdot k_\text{on,g} \, {\left.u_\text{Ga}\right|_\mathcal{S}} + \\ & \qquad+ k_\text{off,g} \, u_\text{g} \, , \\
D_\text{G} \, \hat{\mathbf{n}}_\mathcal{S} \cdot \nabla {\left.u_\text{G}\right|_\mathcal{S}} &= 0\, .
\end{align*}
\end{linenomath}
For the initial state of the system, we assume that all optogenetic GEF is in its inactive form $u_\text{G}$.

\paragraph{Reaction-diffusion dynamics for the Rho module.}
The Rho dynamics are adapted from a previously suggested model \cite{Wigbers.etal2021}.
Rho proteins are assumed to prevail in the cytosol in a GDP-bound state only ($u_\text{R}$) from where it binds to the membrane at a rate $k_\text{on,r}$.
Membrane-bound Rho-GDP ($u_\text{rd}$) unbinds from the membrane at a rate $k_\text{off,R}$.
Nucleotide exchange of Rho-GDP on the membrane is catalyzed at rate $k_\text{r}$ by both native Ect2 on the membrane, as well as by other GEFs ($u_0$) that are not modeled explicitly and are assumed to be homogeneous on the membrane.
The total native GEF concentration is denoted by $u_\text{e0} = u_\text{e} +u_0$.
Optogenetic GEFs ($u_\text{g}$) contribute to the Rho nucleotide exchange with a rate $k_\text{rg}$.
Apart from linear Rho activation, membrane-bound Rho-GTP ($u_\text{rt}$) is also assumed to further enhance Rho activation~\cite{Wigbers.etal2021} via a recruitment term $k_\text{dt}$ that represents the effective autocatalytic interaction between Rho-GTP and its GEFs.
Upon hydrolysis with rate $k_\text{gap}$, Rho-GTP detaches from the membrane.
The corresponding mass-conserving reaction-diffusion equations read
\begin{linenomath}
\begin{align*}
  \partial_t u_\text{R} &= D_\text R \nabla^2 u_\text{R} \, , \\
  \partial_t u_\text{rd} &= D_\text{rd} \nabla_\mathcal{S}^2 u_\text{rd} - \nabla \cdot \bigl(\mathbf v \, u_\text{rd}\bigr) + \\ & \qquad+ f_\text{rd} \bigl({\left.u_\text{R}\right|_\mathcal{S}}, u_\text{rd}, u_\text{rt} \bigr) \, , \\
  \partial_t u_\text{rt} &= D_\text{rt} \nabla_\mathcal{S}^2 u_\text{rt} - \nabla \cdot \bigl(\mathbf v \, u_\text{rt}\bigr) + \\ & \qquad+ f_\text{rt} \bigl({\left.u_\text{R}\right|_\mathcal{S}}, u_\text{rd}, u_\text{rt} \bigr) \, , 
\end{align*}
\end{linenomath}
with reaction terms
\begin{linenomath}
\begin{align*}
  f_\text{rd} \bigl({\left.u_\text{R}\right|_\mathcal{S}}, u_\text{rd}, u_\text{rt} \bigr) &= k_\text{on,r} \, {\left.u_\text{R}\right|_\mathcal{S}} - k_\text{off,R} \, u_\text{rd} + \\
  &  -u_\text{e0} \cdot (k_\text{r} + k_\text{dt} \, u_\text{rt}^2 ) \, u_\text{rd} + \\ & - u_\text{g} \cdot k_\text{rg} \, u_\text{rd} \, , \\[.5em]
  f_\text{rt} \bigl({\left.u_\text{R}\right|_\mathcal{S}}, u_\text{rd}, u_\text{rt} \bigr) &= u_\text{e0} \cdot (k_\text{r} + k_\text{dt} \, u_\text{rt}^2 ) \, u_\text{rd} + \\ & + u_\text{g} \cdot k_\text{rg} \, u_\text{rd} - k_\text{gap} \, u_\text{rt} \, .
\end{align*}
\end{linenomath}
The cytosolic Rho dynamics are subject to the boundary condition
\begin{linenomath}
\begin{align*}
D_\text{R} \, \hat{\mathbf{n}}_\mathcal{S} \cdot \nabla {\left.u_\text{R}\right|_\mathcal{S}} &= - k_\text{on,r} \, {\left.u_\text{R}\right|_\mathcal{S}} + k_\text{off,R} \, u_\text{rd} + \\ & \qquad+ k_\text{gap} \, u_\text{rt} \, .
\end{align*}
\end{linenomath}

\subsection{Mechanical modeling}%
In this section, we specify the model equations and parameters for the mechanical module. For a detailed description of the model, we refer to the Supplementary Information.

Unless specified otherwise, the oocyte membrane mechanics are modeled separately from the chemical modules.
We derive the dynamic equation for the oocyte shape, parameterized in polar coordinates by the oocyte radius $r(\theta)$, from an extended Canham-Helfrich-type free energy functional \cite{Canham1970, Helfrich1973}
\begin{widetext}
\begin{equation*}
    \mathcal H = \! \int \! \mathrm{d}\theta \, \sqrt{g} \, \left[\lambda \, \zeta(u_\text{rt}) {+} \frac{\kappa}{2} \left(\frac{r^2{+}2 \, (\partial_\theta r)^2{-}r \, \partial_\theta^2 r}{\sqrt{g}^3} - H_0 \right)^2 \right] .
\end{equation*}
\end{widetext}
Actomyosin contractility, triggered by the presence of Rho-GTP $u_\text{rt}$, is modeled via a concentration-dependent surface tension-like term $\chi(u_\text{rt})$ with strength $\lambda$\, \cite{Nishikawa.etal2017}.
Deviations from a preferred curvature $H_0$ are counteracted by the bending rigidity $\kappa$.
In the two-dimensional cross-sectional plane, the metric $g{=}r^2 {+}(\partial_\theta r)^2$ accounts for the arc length of the bounding oocyte membrane curve in polar coordinates.
The temporal evolution of the oocyte shape is primarily determined by a relaxation towards the minimum of the effective free energy,
\begin{linenomath}
\begin{equation*}
    \partial_t r = - \frac{1}{\tau} \frac{\delta \mathcal H}{\delta r} \, ,
\end{equation*}
\end{linenomath}
where $\tau$ is the relaxation time scale.
For numerical solutions of the shape dynamics, we additionally account for volume conservation in the time evolution of $r(\theta)$ by including a corresponding Lagrangian multiplier.

\subsection{Simulations}%
The simulations of the reaction-diffusion equations of the chemical modules, uncoupled from the mechanical module, employ finite element methods ({COMSOL Multiphysics~6.0}).
For this, we emulated the cytosol as a 2D domain (circular geometry) and the membrane as the 1D surface of this domain, with parameters as in the Supplementary Tables.
The numerical simulations of the mechanical module are implemented in {Wolfram Mathematica~13.1} using finite element methods, where the results from the simulations of the chemical module were used to prescribe the Rho-GTP concentration dynamics.
For the mechanical module, we used a circular 1D domain corresponding to the oocyte membrane, with parameters as in the Supplementary Tables.
In addition, we performed fully coupled finite element simulations of the chemical and mechanical modules in {Wolfram Mathematica~13.1} (Supplementary Information).
In these simulations, we again used a circular 1D domain representing the membrane, with effective cytosolic dynamics mapped to the membrane \cite{Ziepke.etal2016}.
For parameters, see Supplementary Tables.

\subsection{Excitation threshold}%
To understand the different origins of traveling Rho peaks, we introduced the concept of an excitation threshold $\bar u_\text{rt}$.
This quantity is used to classify the local Rho dynamics and, in particular, identify regions where the Rho dynamics are excitable.
The excitation threshold can be formally defined for a well-mixed system (Supplementary Information).
For a given system state, characterized by the Rho concentrations $u_\text{R}$, $u_\text{rd}$ and $u_\text{rt}$, the excitation threshold is the Rho-GTP concentration at which the change in Rho-GTP concentration $\partial_t u_\text{rt}$ changes sign:
\begin{linenomath}
\begin{align*}
    & f_\text{rt}(u_\text{rd}, u_\text{rt}{=}\bar u_\text{rt}) = 0 \, , \\
    & \left.\frac{\partial f_\text{rt}(u_\text{rd}, u_\text{rt})}{\partial u_\text{rt}} \right|_{\bar u_\text{rt}} > 0 \,  .    
\end{align*}
\end{linenomath}
If this equation has no physically relevant solution, the excitation threshold is taken to be $\bar u_\text{rt} = 0$.
Following this definition, the system is (locally) on an excursion loop in phase space if $u_\text{rt} > \bar u_\text{rt}$, and it is generally relaxing towards a fixed point concentration $u_\text{rt}^*$ if $u_\text{rt} < \bar u_\text{rt}$.
If the fixed point concentration is larger than the smallest possible excitation threshold, the Rho dynamics are oscillatory.
Note that the shape of the nullclines and therefore also the excitation threshold $\bar u_\text{rt}$ depend crucially on the local GEF concentration $u_\text{e0}$.

\newpage

\setcounter{figure}{0} 
\renewcommand{\figurename}{EXTENDED DATA FIG.}

\begin{figure*}
\centering
	\includegraphics[trim={0 9cm 0 0},clip,width=\textwidth]{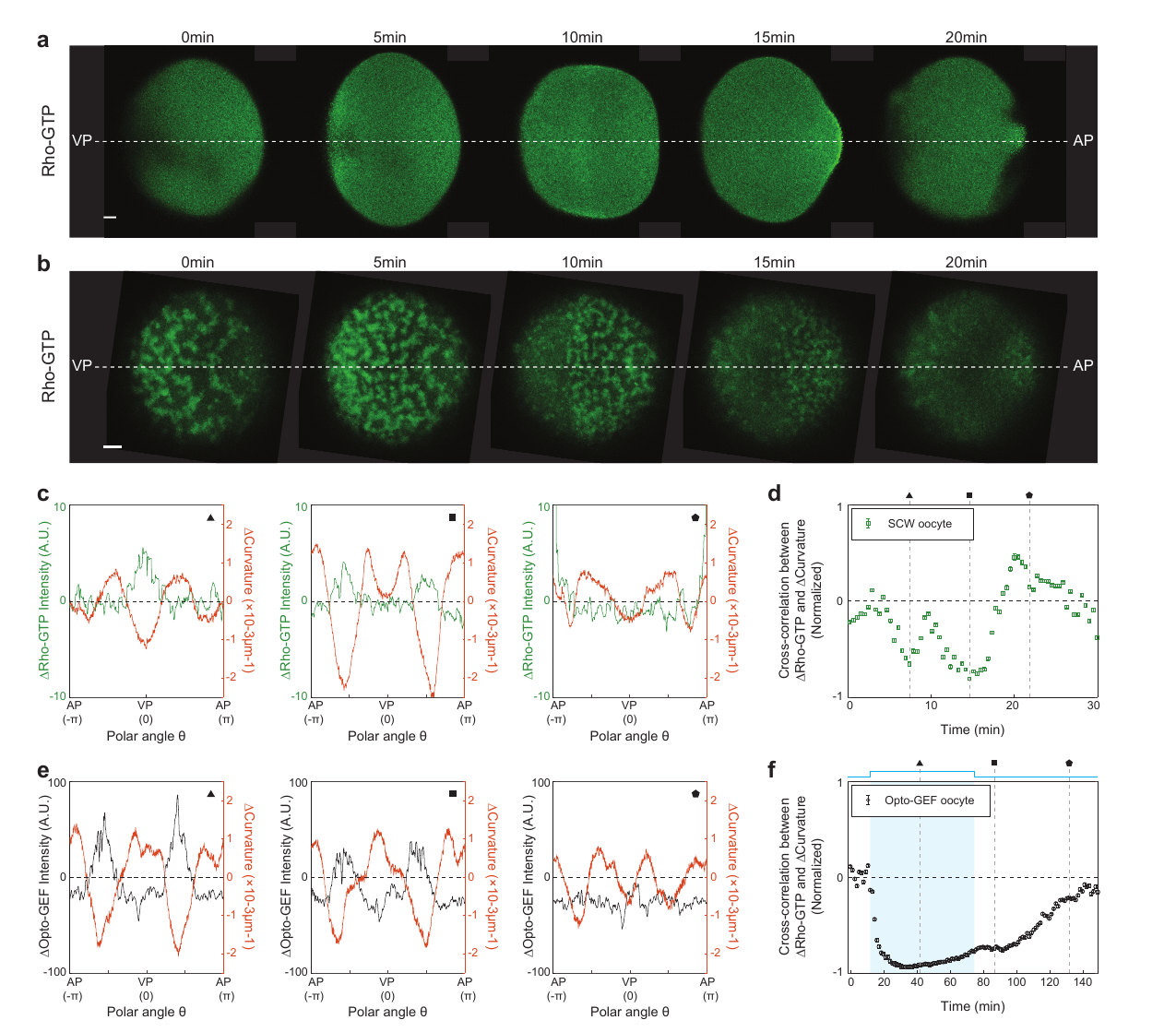}
	\caption{\footnotesize\textbf{Chemo-mechanical activation of starfish oocytes under meiotic and optogenetic GEF signal inputs.}
        \textbf{a,}~Membrane time-lapse fluorescence images showing the propagation of an active Rho band (label: rGBD-GFP) from vegetal pole (AP) to animal pole (AP) during the meiotic anaphase. The Rho-GTP band follows a propagating GEF wavefront (not shown here) in this wild-type case and instructs a subsequent surface contraction wave (SCW). Scale bar: $20\,\mu \text{m}$.
        \textbf{b,}~Membrane time-lapse fluorescence images showing the modulation of Rho cortical excitability by the propagating GEF wavefront during the meiotic anaphase of a GEF-overexpressing oocyte. Oocyte was embedded in 1.5\% low-melt agarose gel to restrict mechanical deformation and retrieve cleanly the biochemical signals. (For a complete discussion of Rho-GTP spiral-formation mechanisms see prior works~\cite{Tan.etal2020,Wigbers.etal2021,Liu.etal2021}). Scale bar: $20\,\mu \text{m}$.
        \textbf{c,}~Distribution of Rho-GTP intensity and membrane curvature changes as a function of oocyte polar angle for the representative snapshots shown in Fig.~\ref{fig1}a-b.
        \textbf{d,}~Cross-correlation score between Rho-GTP intensity and membrane curvature changes during the meiotic time series shown in Fig.~\ref{fig1}a-b. Error bars show the standard error across all membrane sites. The membrane curvature closely follows Rho-GTP dynamics during the onset and propagation phases of the SCW.
        \textbf{e,}~Distribution of Opto-GEF intensity and membrane curvature changes as a function of oocyte polar angle for the representative snapshots shown in Fig.~\ref{fig1}e-f.
        \textbf{f,}~Cross-correlation score between Opto-GEF intensity and membrane curvature changes during the optogenetically-induced time series shown in Fig.~\ref{fig1}e-f. Error bars show the standard error across all membrane sites. The membrane curvature closely follows Opto-GEF dynamics and poses a good proxy for Rho-GTP dynamics in the optogenetic assay.
	}
	\label{ExtFig1}
\end{figure*}

\newpage

\begin{figure*}
\centering
	\includegraphics[trim={0 8.5cm 0 0},clip,width=1\textwidth]{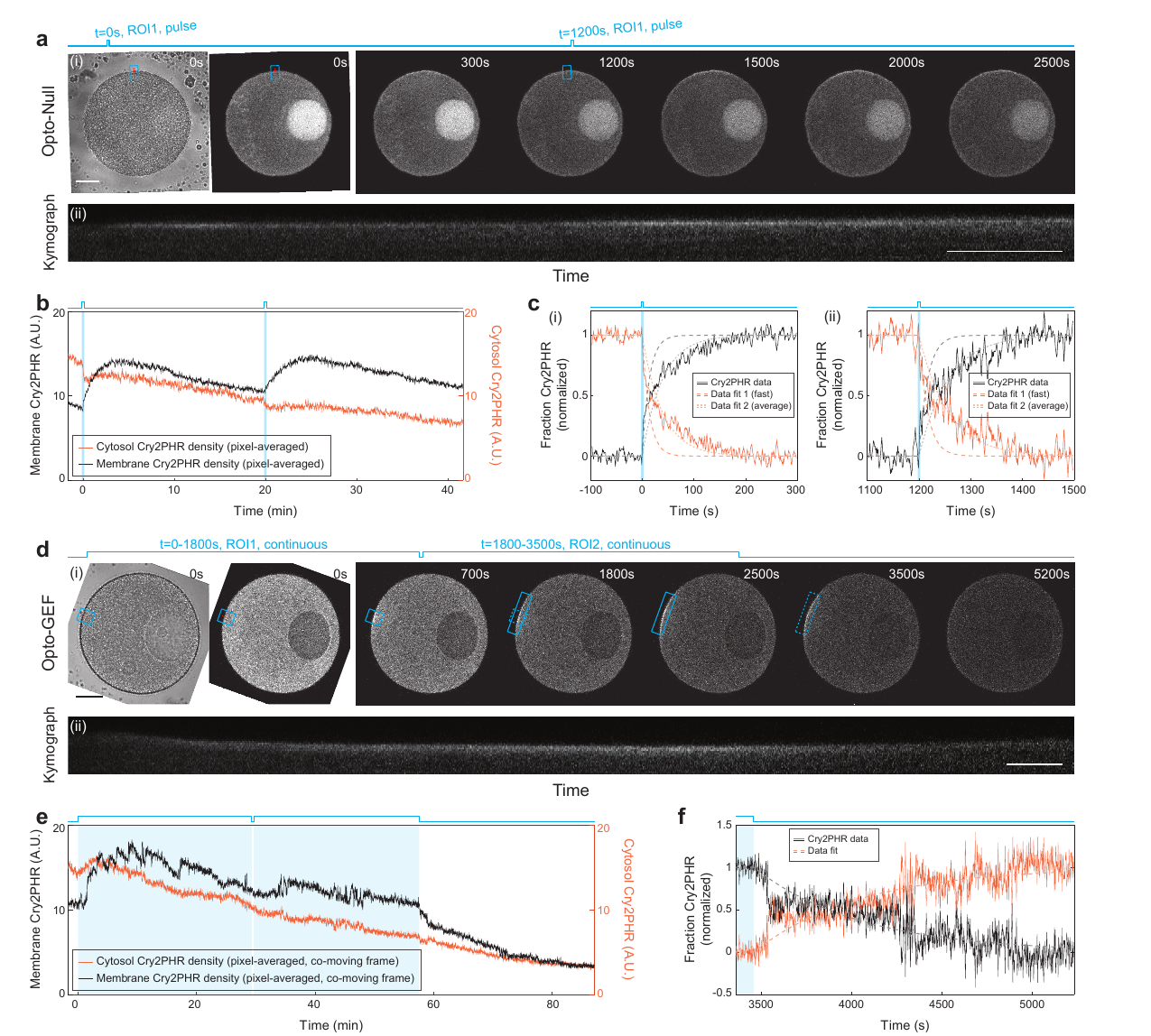}
	\caption{\footnotesize\textbf{The Cry2PHR-CIBN optogenetic switch exhibits signature fast-on-slow-off kinetics in prophase-arrested starfish oocytes.}
        \textbf{a,}~Time-lapse oocyte snapshots (i) and Cry2PHR intensity kymograph (ii) of a representative local pulse-activation experiment using the Opto-Null assay. The regional illumination (488~nm) was made at the same region of interest (ROI, cyan boxes) at t=0~s and t=1200~s and lasted for 1s each. Red line in (i) shows the position chosen to perform the kymograph visualization of Cry2PHR intensity for (ii). Scale bars: 50$\mu \text{m}$ (i), 5 min (ii). 
        \textbf{b,}~Compartmental Cry2PHR intensity quantified as a function of time for the experiment in \textbf{a}. The recruitment of Cry2PHR between cytosol and membrane compartments was extracted using the cytosol (orange box) and membrane (white box) selections shown in \textbf{a}(i).
        \textbf{c,}~Fitting of the post-illumination recruitment data from \textbf{b} gives consistent estimates of the kinetic timescale for the Cry2PHR-CIBN binding to take place (same as Fig.~\ref{fig1}d): $\tau_\text{on,1}=9.4s$, $\tau_\text{on,2}=11.6s$ (fitting the fast regional response) and $\tau_\text{on,1}=26.8s$, $\tau_\text{on,2}=38.7s$ (fitting the average regional response).
        \textbf{d,}~Time-lapse oocyte snapshots (i) and Cry2PHR intensity kymograph (ii) of a representative local continuous-activation experiment using the Opto-Ect2 assay. The regional illumination (488~nm) was made at a smaller ROI at first (t=0-1800~s) and then switched to a larger ROI (t=1800-3500~s, cyan boxes). Red line in (i) shows the position chosen to perform the kymograph visualization of Cry2PHR intensity for (ii) where a significant regional-contractile response can be identified. Scale bars: 50$\mu \text{m}$ (i), 5 min (ii).
        \textbf{e,}~Compartmental Cry2PHR intensity quantified as a function of time for the experiment in \textbf{d}.
        \textbf{f,}~Fitting of the post-removal-of-illumination recruitment data from \textbf{e} gives an estimate of the kinetic timescale for the Cry2PHR-CIBN unbinding to take place: $\tau_\text{off}=411.4s$. We found that the binding and unbinding timescales extracted from \textbf{c} and \textbf{f} are consistent throughout all optogenetic experiments performed in this study.
	}
	\label{ExtFig2}
\end{figure*}

\newpage

\begin{figure*}
\centering
	\includegraphics[width=1\textwidth]{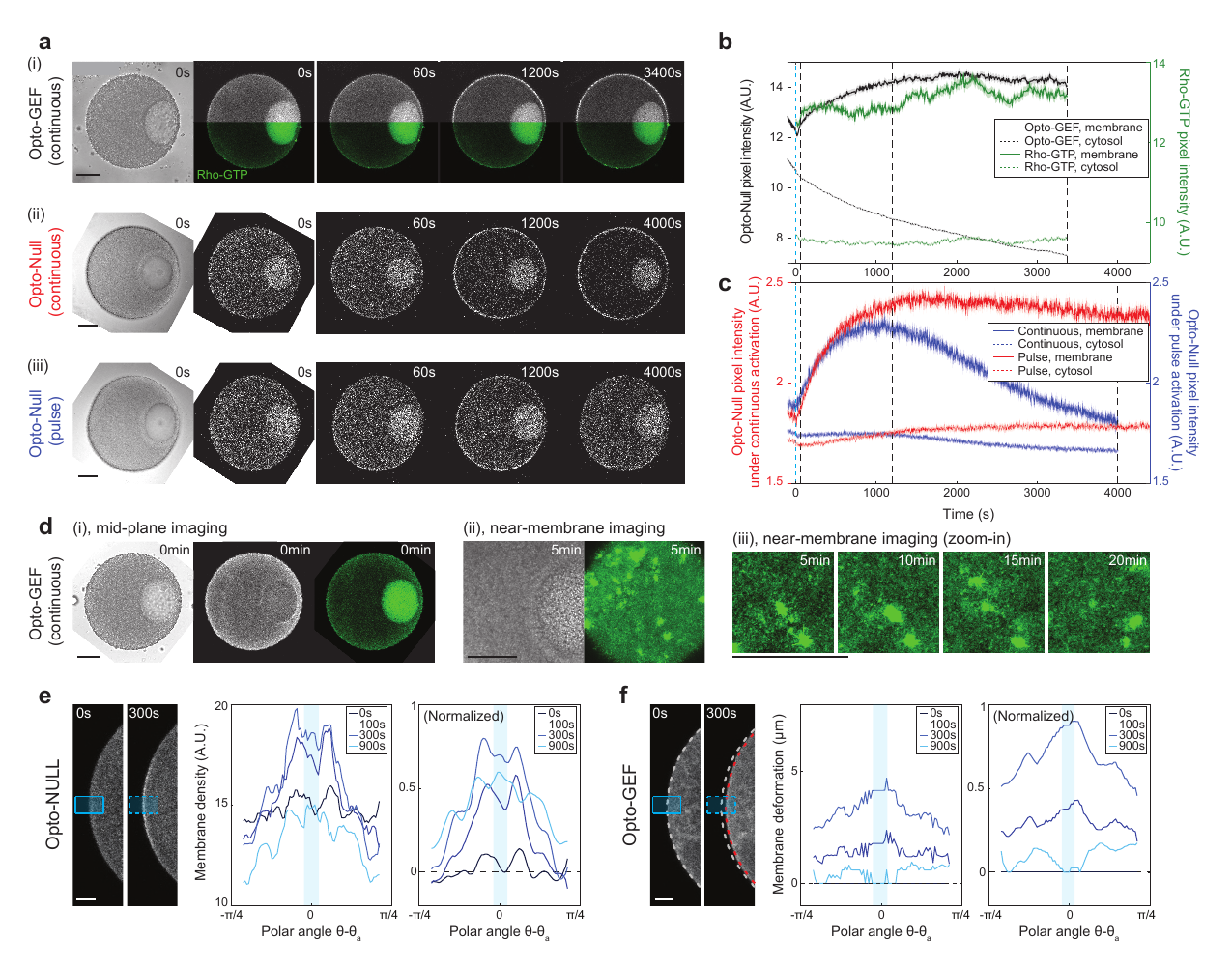}
	\caption{\footnotesize\textbf{Quantification of light-induced chemo-mechanical kinetics from optogenetic assays.}
        \textbf{a,}~Time-lapse oocyte snapshots for representative experiments of (i) continuous global activation of Opto-GEF assay with active Rho markers (Also see Movie S2), (ii) continuous global activation of Opto-Null assay and (iii) pulse global activation of Opto-Null assay. Scale bar: 50$\mu \text{m}$.
        \textbf{b,}~Compartmental Cry2PHR and Rho-GTP intensity quantified as a function of time for the experiment shown in \textbf{a}(i). Shaded area represents standard error across all identified membrane or cytosol sites. The recruitment kinetics were then normalized and plotted as Fig.~\ref{fig2}d.
        \textbf{c,}~Compartmental Cry2PHR intensity quantified as a function of time for the experiments shown in \textbf{a}(ii) and (iii). Shaded area represents standard error across all identified membrane or cytosol sites. The recruitment kinetics were normalized and plotted as Fig.~\ref{fig2}b.
        \textbf{d,}~Time-lapse oocyte snapshots for an representative experiment of continuous global activation of Opto-GEF assay with active Rho markers, where the activation and imaging plane was moved from middle (i) to bottom (ii) of the chamber to capture active Rho dynamics near membrane field of view (see also Movie S3). Panel (iii) shows a zoom-in view of time series (ii), where oscillatory Rho signals show up. Scale bar: 50$\mu \text{m}$.
        \textbf{e,}~Time-lapse edge snapshots of an representative oocyte expressing the Opto-Null assay under a regional pulse (1s) illumination applied at t=0s (Left) and corresponding membrane recruitment kinetics quantified for the photosensitive protein tag complex (Cry2PHR). Scale bar: 20$\mu \text{m}$.
        \textbf{f,}~Time-lapse edge snapshots of an representative oocyte expressing the Opto-GEF assay under a regional pulse (1s) illumination applied at t=0s (Left) and corresponding membrane deformation kinetics quantified for the maximal contracting edge location. The normalized kinetics were plotted as Fig.~\ref{fig2}c. Scale bar: 20$\mu \text{m}$.
	}
	\label{ExtFig3}
\end{figure*}

\newpage

\begin{figure*}
\centering
	\includegraphics[width=1\textwidth]{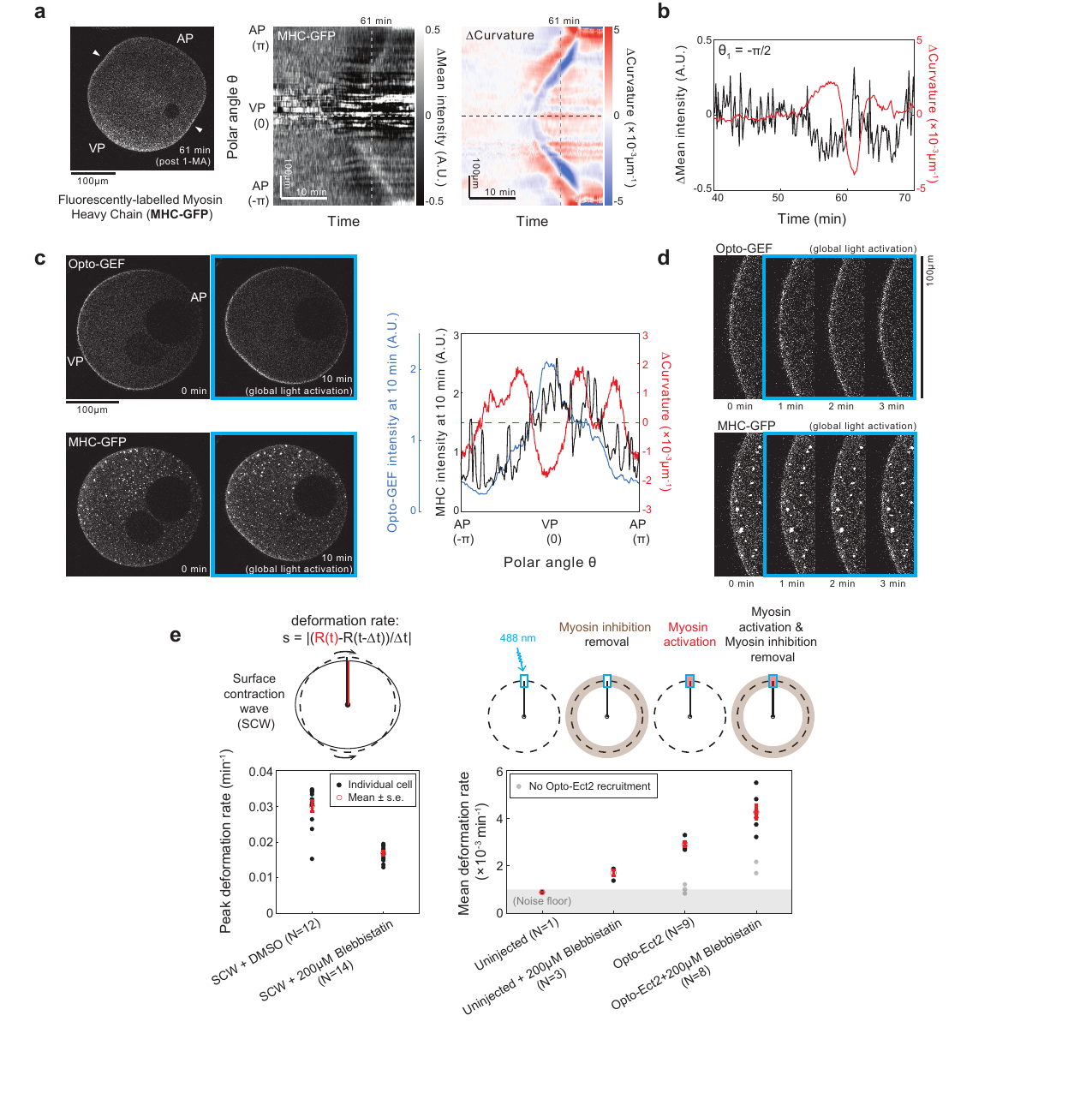}
 \caption{\footnotesize Figure caption on following page.}\label{ExtFig4}
\end{figure*}
\clearpage

\begin{minipage}{\textwidth}\footnotesize
	\textbf{Myosin imaging and functional inhibition for Opto-Ect2 photoactivation.}
        \textbf{a,}~A representative starfish oocyte expressing MHC-GFP undergoes surface contraction wave in meiosis (left). A travelling wave of cortically-recruited myosin (middle kymograph) closely guides the surface curvature changes (right kymograph).
        \textbf{b,}~Temporal traces showing the labelled myosin fluorescent intensity and surface curvature for a representative site on the oocyte boundary as the meiotic surface contraction wave passes in panel~\textbf{a}.
        \textbf{c,}~Left: Pre-activation and post-activation snapshots for a representative oocyte loaded with Opto-GEF assay and MHC-GFP. The oocyte was first globally illuminated for 30~s (488~nm), then a second snapshot was taken 10~min after the initial illumination. Right: Near-membrane distribution of post-activation Opto-GEF and MHC intensity signals (left axis) plotted together with the distribution of surface curvature change (right axis) between pre- and post-activation snapshots.
        \textbf{d,}~Representative near-membrane time-lapse images for a representative oocyte loaded with Opto-GEF assay and MHC-GFP. The oocyte was subjected to continuous global illumination (1~s per frame) instead of the short time exposure in panel~\textbf{c}.
        \textbf{e,}~Top left: Schematic for the definition of meiotic oocyte deformation rate.
        Bottom left: For meiotic surface contraction waves, the peak deformation rate evaluated both for oocytes without blebbistatin inhibition (DMSO treatment) and oocytes with 200~$\mu\text{M}$ blebbistatin inhibition.
        Top right: Schematic for the local light illumination (488~nm) performed for 4 groups of oocytes: no Opto-Ect2 expression and no blebbistatin inhibition, no Opto-Ect2 expression and with blebbistatin inhibition, with Opto-Ect2 expression and no blebbistatin inhibition, with Opto-Ect2 expression and with blebbistatin inhibition.
        Bottom right: For light-activation time series, the mean deformation rate for the 4 groups of oocytes. In groups with Opto-Ect2 expression, a subset of oocytes show no Opto-Ect2 membrane recruitment after illumination (related to variable membrane anchor co-expression) and was shown as grey dots. The mean deformation rate inside each group was calculated excluding the no Opto-Ect2-recruitment oocytes. Error bars show the standard error within each group after making this exclusion.
\end{minipage}

\newpage

\begin{figure*}
\centering
	\includegraphics[width=1\textwidth]{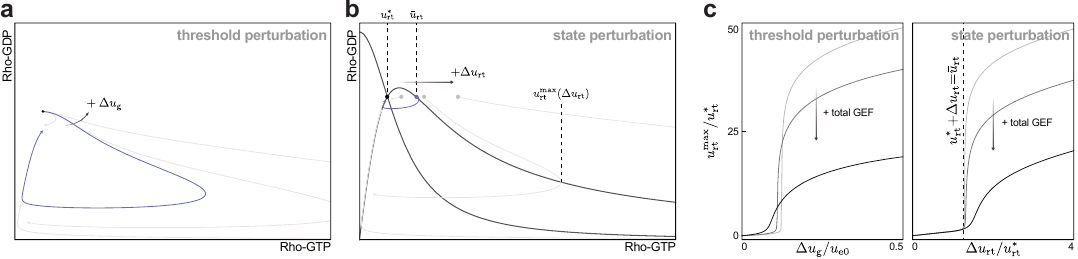}
	\caption{\footnotesize\textbf{Phase portrait analysis of Rho excitation.}
        \textbf{a,}~Phase space trajectories for different membrane GEF increments $\Delta u_\text{g}$ after photoactivation. The blue trajectory corresponds to the smallest GEF increase for which the excitation threshold $\bar u_\text{rt}$ is zero.
        \textbf{b,}~System state and phase space trajectories before (black) and after (gray) a sudden increase $\Delta u_\text{rt}$ in the Rho-GTP concentration. An excitation is caused only after the Rho-GTP concentration crosses the excitation threshold $\bar u_\text{rt}$ (blue). The amplitude of the excitation is the largest Rho-GTP concentration during the excursion, $u_\text{rt}^\text{max}$.
        \textbf{c,}~Amplitude of the excitation $u_\text{rt}^\text{max}$ at varying perturbation strengths for threshold (left, change GEF increase $\Delta u_\text{g}$) and state (right, change Rho-GTP increase $\Delta u_\text{rt}$) perturbations. The excitation amplitude and minimal required perturbation depend on the total GEF concentration on the membrane, most importantly the native GEFs.
	}
	\label{ExtFig5}
\end{figure*}

\newpage

\begin{figure*}
\centering
	\includegraphics[trim={0 11.5cm 0 0},clip,width=1\textwidth]{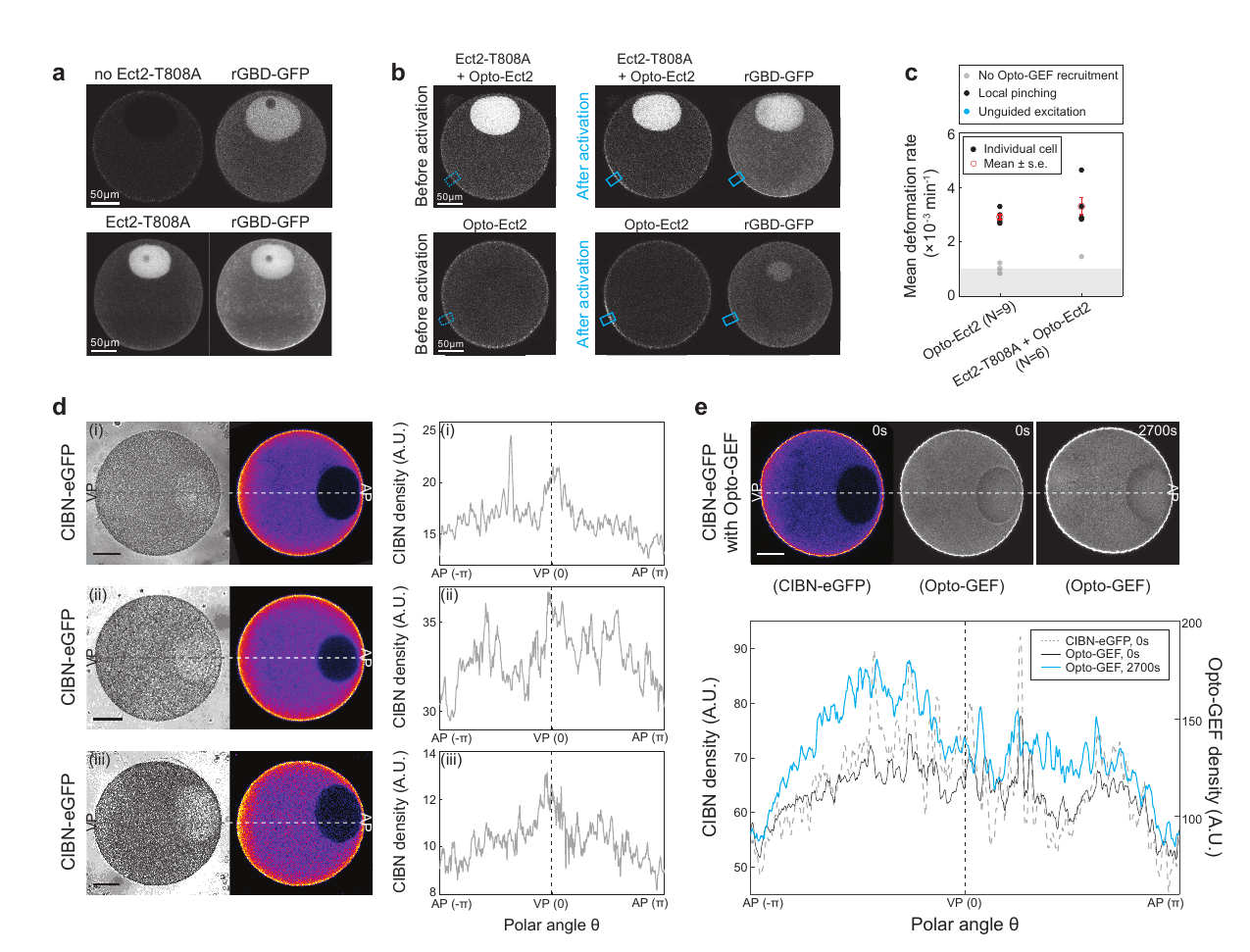}
        \caption{\footnotesize\textbf{Opto-GEF-based oocyte optogenetic responses under local and global illumination.}
        \textbf{a,}~For local illumination of Opto-GEF assay-loaded oocytes, only pinching responses were observed even with additional expression of active endogenous Ect2 (Ect2-T808A). Top: Representative snapshot of a Rho-GTP activity measurement (rGBD-GFP) in oocytes with (top) and without (bottom) Ect2-T808A expression.
        \textbf{b,}~Top: Representative pre- and post-activation snapshots of oocytes co-expressing Ect2-T808A together with the Opto-Ect2 assay. Bottom: Representative pre- and pos-activation snapshots of oocytes expressing only the Opto-Ect2 assay. The oocytes underwent local illumination (cyan box) performed at 1~Hz for 12~min. Note that the different nucleus size in top and bottom rows is due to the different nucleus positions relative to the mid-plane $z$~focus.
        \textbf{c,}~Both the Ect2-T808A co-expressed with the Opto-Ect2 assay and only the Opto-Ect2 assay show comparable mean boundary deformation rates after light activation, and no unguided excitation was observed. The boundary deformation rate was defined and calculated in the same way as in Extended Data Fig.~\ref{ExtFig4}e.
        \textbf{d,}~For global illumination of Opto-GEF assay-loaded oocytes, a nucleus-dictated anchor distribution directs the pole-to-pole asymmetry of Opto-GEF recruitment. Left: Representative snapshots of three oocytes expressing CIBN-eGFP. 
        Right: Quantification of the snapshots shows the nucleus-dictated asymmetry of the anchor distribution on the membrane, that CIBN binding sites are more enriched near VP and less dense near AP. Scale bar: $50\,\mu \text{m}$.
        \textbf{e,}~Top: Representative time-lapse snapshots of a globally-activated oocyte expressing the Opto-GEF assay. 
        Bottom: Quantification of the global activation time series in \textbf{c} shows the asymmetry of the CIBN anchor distribution and the asymmetry-guided accumulation of Opto-GEF on the membrane at long illumination time. Scale bar: $50\,\mu \text{m}$.
    }
    \label{ExtFig6}
\end{figure*}

\newpage

\begin{figure*}
\centering
	\includegraphics[width=1\textwidth]{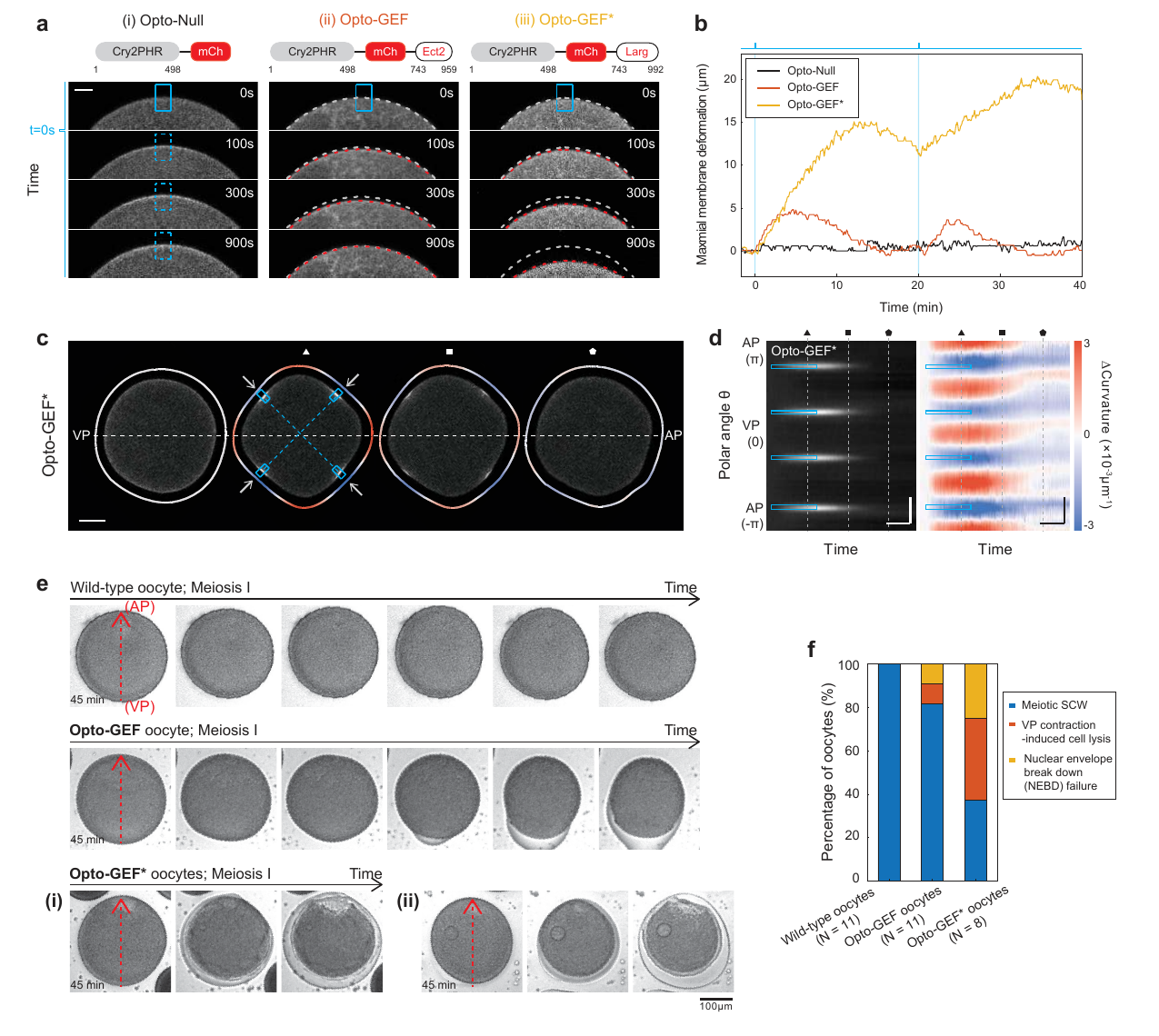}
	\caption{\footnotesize\textbf{Exogenous GEF catalytic domain constituting optogenetic assay (Opto-GEF*) permits strong pinching responses and exhibits a strong ``passive'' catalytic effect.}
        \textbf{a,}~Time-lapse oocyte snapshots in representative local activation experiments for (i) Opto-Null, (ii) Opto-GEF and (iii) Opto-GEF* assays. The regional illumination (488~nm) was made at the same region of interest (ROI, cyan boxes) at t=0~s and t=1200~s and lasted for 1~s each. Scale bar: 20~$\mu \text{m}$.
        \textbf{b,}~Quantification of concurrent maximal edge displacements as a function of time for the experiments shown in \textbf{d}. While the recruitment of Opto-Null does not elicit mechanical responses, the membrane recruitment of Opto-GEF and Opto-GEF* are both capable of inducing edge contractile responses that are reversible upon the removal of illumination. Compared to Opto-GEF, Opto-GEF* assay exhibits stronger mechanical response per unit recruitment of Cry2PHR fluorescence.
        \textbf{c,}~Time-lapse snapshots of a prophase-arrested oocyte expressing Opto-GEF* undergoing reversible shape deformation driven by patterned illumination. The illumination sites are shown in cyan boxes and the ring overlay visualizes the change of cell surface curvatures. Scale bar: 50~$\mu \text{m}$.
        \textbf{d,}~Boundary kymographs of Opto-GEF* density and cell surface curvature change in the course of illumination. Scale bar: 100~$\mu \text{m}$, 10~min.
        \textbf{e,}~Representative bright field time-lapse images for oocytes undergoing hormone (1-MA) induced meiosis I. At 45~min post the hormone induction, nuclear envelope has broken down, setting up the VP-AP polarity. In wild-type oocytes (upper row), a surface contraction wave (SCW) starts from VP and propagates towards AP within the following hour. For oocytes loaded with Opto-GEF assay (middle row), despite that the SCW starts at VP, an elevated VP contraction causes the oocyte to bleb and the SCW does not propagate. For oocytes loaded with Opto-GEF* assay (bottom row), instead of the onset of SCW at VP, a globally-elevated contraction gradient centered at VP was observed. At long time (variable, 1-2~h post hormone induction), the oocytes undergo lysis which start from AP. Note that in all meiotic oocytes, no light activation was applied throughout. 
        \textbf{f,}~Percentage of oocyte behaviors as from panel~\textbf{e}. The representative behaviors shown in Opto-GEF and Opto-GEF* oocytes from panel~\textbf{e} are categorized as ``VP contraction-induced cell lysis''.
	}
	\label{ExtFig7}
\end{figure*}

\newpage

\begin{figure*}
\centering
	\includegraphics[width=1\textwidth]{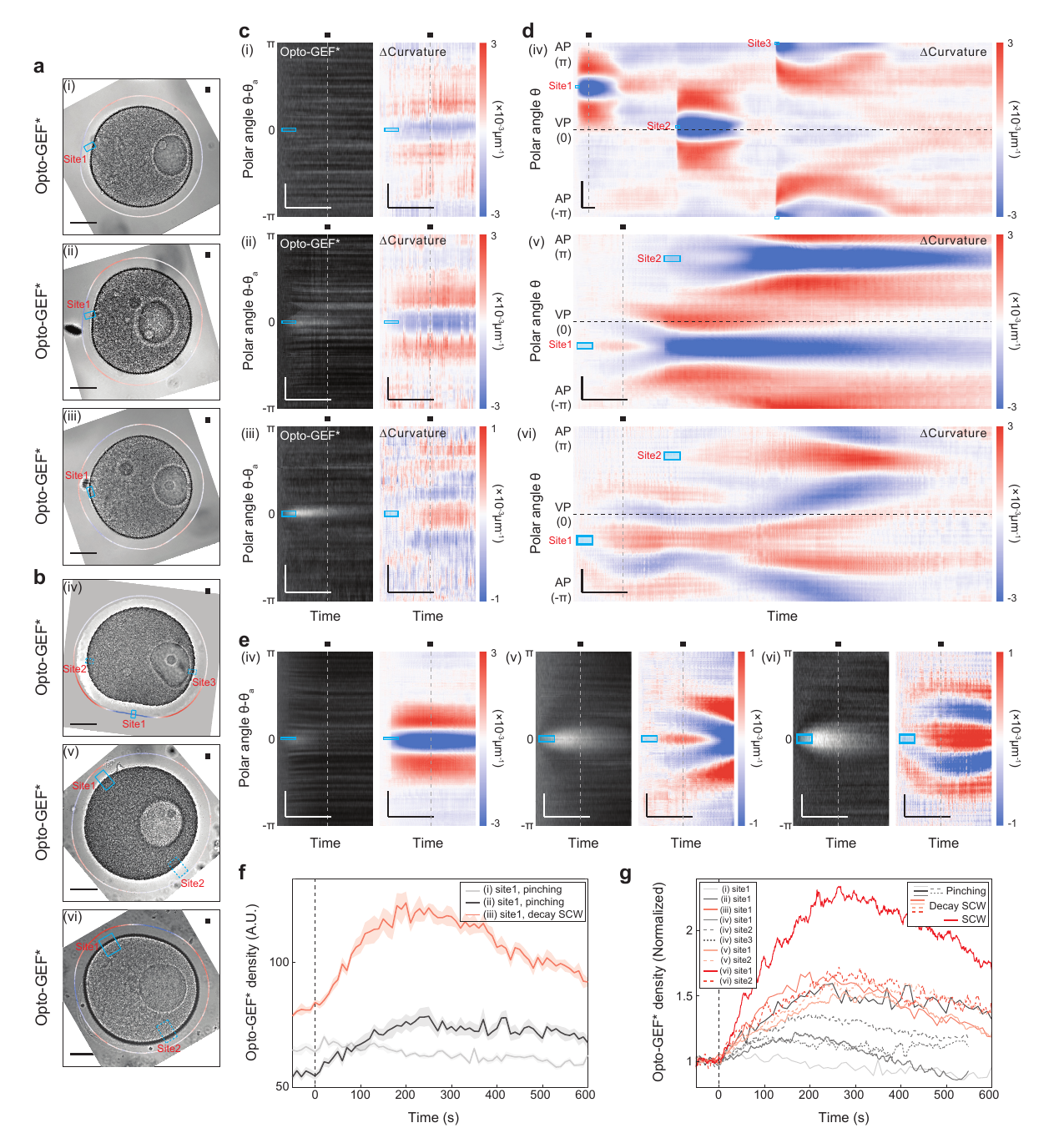}
 \caption{Figure caption on following page.}
 \label{ExtFig8}
\end{figure*}

\clearpage

\begin{minipage}{\textwidth}\footnotesize
    \textbf{Generation and quantification of unguided photoactivated SCWs with Opto-GEF* assay.}
    \textbf{a,}~Brightfield snapshots of locally-activated oocytes expressing (i) low amount (ii) intermediate amount and (iii) high amount of Opto-GEF* constructs collected from same-day experiments. Regional illumination was made in ROIs with consistent sizes (cyan boxes, $\approx 250\,\mu \text{m}^{2}$) near VP and lasted for 180~s (with $0.1\,\text{Hz}$ frequency). Respectively, oocytes (i), (ii) and (iii) show pinching and SCW that later decays when propagating. Snapshots were taken 10~min after start of illumination. Ring overlay visualizes the change of membrane curvature in each case. Scale bar: $50\,\mu \text{m}$.
   \textbf{b,}~Brightfield snapshots of locally-activated oocytes expressing high amounts of Opto-GEF* constructs and activated with (iv) small (v) medium and (vi) large areas of illumination. Regional illumination was made sequentially in multiple ROIs (cyan boxes, site1-3 for oocyte (iv), and site1-2 for oocyte (v) and (vi)) throughout the time series. Each illumination lasted for 200~s with $1\,\text{Hz}$ frequency. Snapshots were taken 10~min after the start of the first round of illumination. Ring overlay visualizes the change of membrane curvature in each case. Scale bar: $50\,\mu \text{m}$. See Supplementary Information for a detailed discussion of the experiments.
   \textbf{c,}~Kymographs of Opto-GEF* intensity and membrane curvature change for locally-activated oocytes (i)-(iii) (shown in \textbf{a}) after illumination. Scale bar: $100\,\mu \text{m}$, 10~min.
   \textbf{d,}~Kymographs of Opto-GEF* intensity and membrane curvature change for locally-activated oocytes (iv)-(vi) (shown in \textbf{b}) through the course of multi-site illumination. Scale bar: $100\,\mu \text{m}$, 10~min.
   \textbf{e,}~Zoom-in kymographs of Opto-GEF* intensity and membrane curvature change for locally-activated oocytes (iv)-(vi) after the first round of illumination. Oocytes (iv) and (vi) were presented as representative cases of pinching and SCW in Fig.~\ref{fig4}b and c. Scale bar: $100\,\mu \text{m}$, 10~min.
   \textbf{f,}~Quantification of Opto-GEF* membrane density accumulation over the photoactivation of oocytes (i)-(iii). The line shows the mean density accumulation inside each illumination site, and the shaded area shows the standard deviation of accumulated density for sites rotated in a range of 10 degrees on oocyte boundaries.
   \textbf{g,}~Quantification of the propagation distances of negative curvature bands over the local activation of oocytes (i)-(iii) and the multiple rounds of local activation for oocytes (iv)-(vi). Data from \textbf{g} together with the quantified propagation distances of negative curvature bands (Methods) for each local activation case were plotted as Fig.~\ref{fig4}e.
\end{minipage}

\newpage

\begin{figure*}
\centering
	\includegraphics[width=1\textwidth]{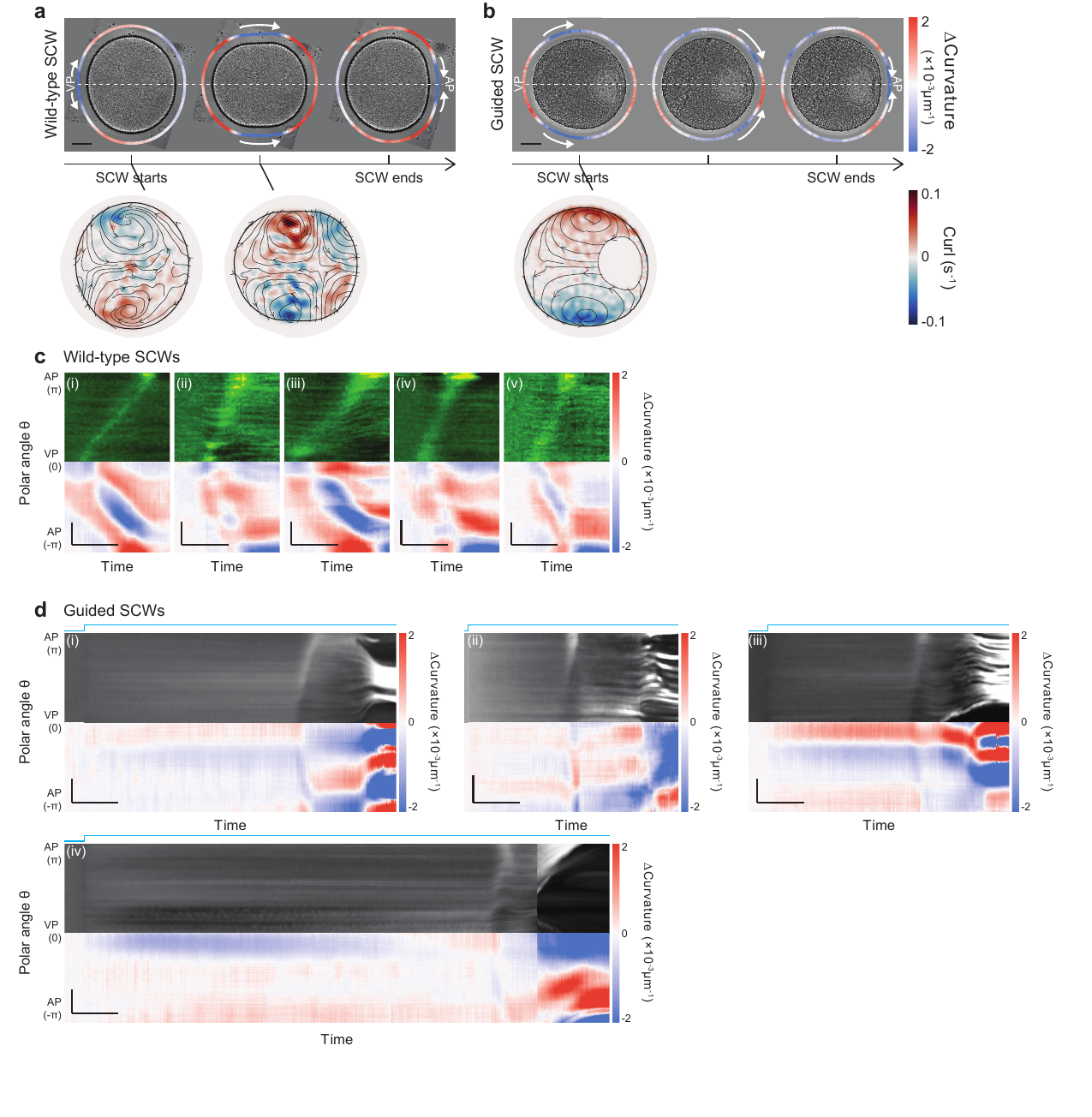}
	\caption{\footnotesize\textbf{Generation and quantification of guided photoactivated SCWs with Opto-GEF and wild-type assays.}
        \textbf{a,}~Cytoplasmic flows and the corresponding curl field visualized at the onset of a representative wild-type oocyte undergoing SCW. The streamline view of cytoplasmic flows was generated from a PIV-extracted velocity field from bright field images corrected for the center-of-mass frame.
        \textbf{b,}~Cytoplasmic flows and the corresponding curl field visualized at the onset of a representative Opto-Ect2 assay-loaded oocyte undergoing global illumination-induced guided SCW.
        \textbf{c,}~Kymographs of Opto-GEF intensity and membrane curvature changes for oocytes exhibiting guided SCWs after continuous global illumination ($N=4$). Scale bar: $100\,\mu \text{m}$, 10~min. The kymographs cover the consistent oocyte responses under global activation, that Opto-GEF first accumulates on the membrane asymmetrically, then this VP-AP gradient induces a fast SCW propagating from VP to AP. After passing of the guided SCW, the oocytes exhibit mechanical instability that leads to eventual cell lysis. See Supplementary Information for a detailed discussion.
        \textbf{d,}~Kymographs of Rho-GTP intensity and membrane curvature changes for wild-type oocytes undergoing SCWs after induction of meiosis ($N=5$). Scale bar: $100\,\mu \text{m}$, 10~min.
	}
	\label{ExtFig9}
\end{figure*}

\newpage

\begin{figure*}
\centering
	\includegraphics[trim={0 10cm 0 0},clip,width=1\textwidth]{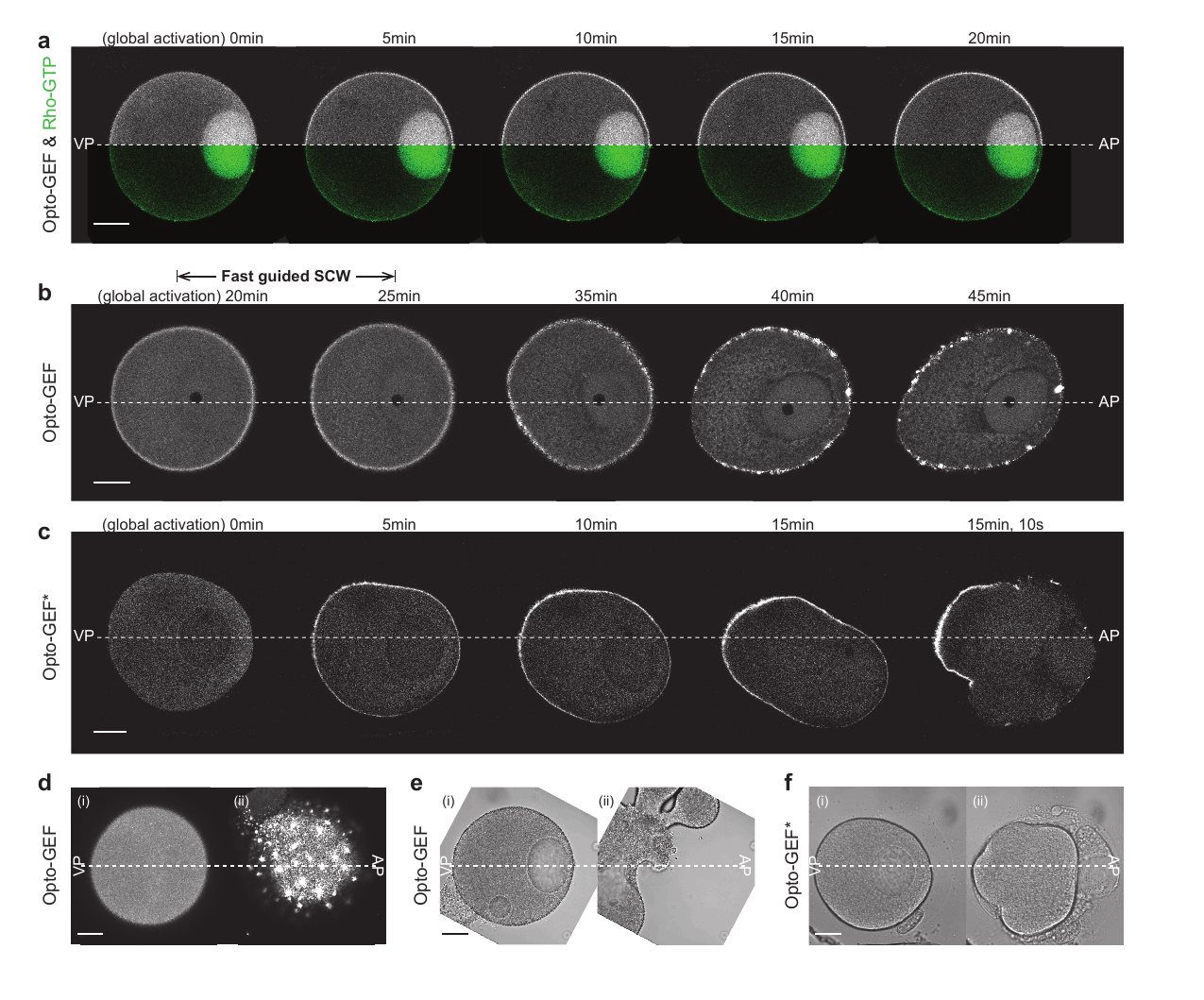}
	\caption{\footnotesize\textbf{Global activation of oocytes with high Opto-GEF or Opto-GEF* expression levels drives cell lysis.}
        \textbf{a,}~Time-lapse fluorescence images of an oocyte expressing Opto-GEF and active Rho marker constructs undergoing continuous global illumination. Scale bar: $50\,\mu \text{m}$.
        \textbf{b,}~Time-lapse fluorescence images of an oocyte expressing Opto-GEF assay undergoing continuous global illumination until the onset of mechanical instability and cell lysis. Scale bar: $50\,\mu \text{m}$.
        \textbf{c,}~Time-lapse fluorescence images of an oocyte expressing Opto-GEF* assay undergoing continuous global illumination until the onset of mechanical instability and cell lysis. Scale bar: $50\,\mu \text{m}$.
        \textbf{d,}~Representative near-membrane fluorescence snapshots for oocytes expressing the Opto-GEF assay (i) before and (ii) after the illumination-induced lysis. Scale bar: $50\,\mu \text{m}$. The appearance of regularly-spaced asters near the membrane is consistent for all Opto-GEF-induced lysis events.
        \textbf{e,}~Representative bright field snapshots for oocytes expressing the Opto-GEF assay (i) before and (ii) after the illumination-induced lysis. Scale bar: $50\,\mu \text{m}$. Oocytes placed in this condition undergo drastic bursting lysis immediately following the formation of membrane asters.
        \textbf{f,}~Representative bright field snapshots for oocytes expressing the Opto-GEF* assay (i) before and (ii) after the illumination-induced lysis. Scale bar: $50\,\mu \text{m}$. Oocytes in this condition undergo gradual compressional changes that eventually lead to the nucleus being omitted as the heterogeneous structure. See Supplementary Information for detailed discussions.
	}
	\label{ExtFig10}
\end{figure*}

\end{document}